\documentclass[letterpaper]{aa}
 
\usepackage{graphicx}
\usepackage{amsmath,amsfonts}
 
\usepackage{natbib} 
\bibpunct{(}{)}{;}{a}{}{,} 

\let\sun=\odot
\newcommand {\kms}{\ensuremath{\mathrm{km\,s}^{-1}}}

\newcommand{\modif}[1]{#1} 

\begin{document}

  \title{A new Monte Carlo code for star cluster simulations: II.~Central black hole and stellar collisions}
 
  \author{M. Freitag\inst{1,2}
  \and W. Benz\inst{2}}
 
\institute{
  California Institute of Technology, Mail code 130-33, Pasadena, CA 91125, USA
  \and Physikalisches Institut,
  Universit\"at Bern,
  Sidlerstrasse 5,
  CH-3012 Bern,
  Switzerland}
 
\offprints{M. Freitag}
\mail{freitag@tapir.caltech.edu}

\date{Received / Accepted}
 
\titlerunning{Monte Carlo Cluster Simulations~II.}
\authorrunning{M. Freitag \& W. Benz}

\abstract{ We have recently written a new code to simulate the long
  term evolution of spherical clusters of stars. It is based on the
  pioneering Monte Carlo scheme proposed by H\'{e}non in the 70's.
  Unlike other implementations of this numerical method which were
  successfully used to investigate the dynamics of globular clusters,
  our code has been devised in the specific goal to treat dense
  galactic nuclei. In a previous paper, we described the basic version
  of our code which includes 2-body relaxation as the only physical
  process. In the present work, we go on and include further physical
  ingredients that are mostly pertinent to galactic nuclei, namely the
  presence of a central (growing) black hole (BH) and collisions
  between (main sequence) stars. Stars that venture too close to the
  BH are destroyed by the tidal field. We took particular care of this
  process because of its importance, both as a channel to feed the BH
  and a way to produce accretion flares from otherwise quiescent
  galactic nuclei. Collisions between stars have often been proposed
  as another mechanism to drive stellar matter into the central
  BH. Furthermore, non disruptive collisions may create peculiar
  stellar populations which are of great observational interest in the
  case of the central cluster of our Galaxy. To get the best handle on
  the role of this process in galactic nuclei, we include it with
  unpreceded realism through the use of a set of more than 10\,000
  collision simulations carried out with a SPH (Smoothed Particle
  Hydrodynamics) code. Stellar evolution has also been introduced in a
  simple way, similar to what has been done in previous dynamical
  simulations of galactic nuclei. To ensure that this physics is
  correctly simulated, we realized a variety of tests whose results
  are reported here. This unique code, featuring most important
  physical processes, allows million particle simulations, spanning a
  Hubble time, in a few CPU days on standard personal computers and
  provides a wealth of data only rivalized by $N$-body simulations. 
   }
 
  \maketitle
 
\section{Introduction}
\label{sec:intro}

This paper is the second part of the description of the code
we have developed in the past few years in order to
investigate the long-term dynamics of dense galactic
nuclei. In a first paper \citep[ hereafter paper~I]{FB01a},
we presented the basic version of this Monte Carlo (MC) code
which deals with 2-body relaxation. In this article, we add
flesh to this kernel by incorporating physical
effects that are of particular interest and relevance for
galactic nuclei.

The structure of the paper is as follows. In Sec.~\ref{subsec:motiv},
we motivate our interest in the dynamics of galactic nuclei through a
short review of the history of the study of this field. We then
proceed to describe the new physics incorporated in the code. The
principles of the basic version of our MC code are presented in
paper~I with which the reader is advised to get familiar; we don't
repeat this information here. Our treatment of stellar collisions is
treated in Sec.~\ref{sec:coll}, tidal disruptions in
Sec.~\ref{sec:disr}, while further, more minor, additions and
improvements are described in Sec.~\ref{sec:imp}. A variety of test
simulations are reported and discussed in
Sec.~\ref{sec:tests}. Finally we summarize this work and propose
future developments in Sec.~\ref{sec:concl}. An appendix is added to
expose how we build initial conditions for use with our code.

\subsection{Astrophysical motivation}
\label{subsec:motiv}

Only very few reviews have been written about the dynamics of galactic
nuclei (\citealt{Gerhard94} is the only recent reference known to us),
hence, we introduce our work by a summary of the history of this
complex field.

The theoretical study of the dynamics of galactic nuclei was initiated
in the 60's, mostly to investigate whether stellar collisions in
extremely dense clusters could power the, then recently discovered,
quasars (QSOs). In these early speculations, the presence of a central
massive black hole (MBH) wasn't assumed. QSO observations required the
concentration of a huge amount of matter in a small volume and it was
thought that a stellar system could expel its angular momentum and
contract more easily than a purely gaseous configuration, hence
reaching densities such that highly energetic collisions between stars
should be commonplace \citep{GAR65,VonHoerner68}. While
\citet{Woltjer64} proposed that collisions themselves would be a
strong source of optical radiation and radio-emitting energetic
particles, others pointed out that these disruptive events should lead
to the formation of a massive compact gas object \citep{UW64},
maintained at QSO luminosity either by further collisions
\citep{SS66,SS67} or massive star formation and supernovae (SN) explosions 
\citep{Sanders70b}. Growth of high-mass stars through repeated
mergers in a low-velocity collisional cluster was proposed by
\citet{Colgate67} as another way of forming SN-powered QSOs.
Unfortunately, in none of these
early studies, was the stellar dynamics treated in a realistic way,
most authors having recourse to some extension of the evaporative
model of globular clusters \citep[see, e.g.,][]{Spitzer87}. In
particular, the process of gravothermal collapse was not known and the
role of mass segregation not properly recognized.

Nearly all further studies accounted for the presence of a central
MBH, an object more and more widely accepted as necessary to explain
QSOs and others Active Galactic Nuclei (AGN), and likely to be present
in at least some normal present-day nuclei, as a relic of past
activity \citep{LyndenBell69}. In a relaxed cluster were stars are
destroyed in the vicinity of the BH, presumably by tidal forces
\citep{Hills75}, their (quasi-)stationary density distribution was predicted
to be a power-law, $n_\ast \propto R^{-7/4}$
\citep{Peebles72,BW76,BW77}. The tidal disruption rate is dominated by
stars that are brought onto very elongated orbits by relaxation
\citep[][ see Sec.~\protect\ref{sec:disr_lc}]{FR76,LS77,YSW77,CK78}.
This rate is probably insufficient to feed a QSO-class MBH unless the
stellar density is so high that collisional gas production should
dominate
\citep{YSW77,Young77b,SW78,Frank78}. The link between the earlier BH-free 
stellar dynamical models for AGN and these studies of MBH-cluster
systems was traced by
\citet{BR78} who showed that most very dense stellar systems will
naturally evolve to form large BHs.

In the 80's, self-consistent simulations of the evolution of galactic
nuclei appeared, that confirmed these conclusions
\citep{McMLC81,DS83}. These models were based on Fokker-Planck and
Monte Carlo codes first developed to study globular clusters. A
serious shortcoming of these works was to assume that collisions were
completely disruptive. \citep{DDC87a,DDC87b} and
\citet{MCD91} improved on this by implementing some extension of the
simple semi-analytical prescription of \citet{SS66} to account for
partial disruptions but the introduction of collisions into
Fokker-Planck codes had to be done in a quite unrealistic way (see
Sec.~\protect\ref{subsec:collgalmod}).
Stellar evolution was also
included with the conclusion that, provided a significant fraction
of the emitted gas is accreted, it dominates the feeding of the BH in
systems of moderate stellar density while collisions are still the
main player in denser nuclei and that the full range of AGN and QSO
luminosities can be attained without having recourse to an external
source of gas. More recently, \citet{Rauch99} has considered the
relativistic dynamics of a compact stellar cluster dominated by a
central MBH in an AGN and concluded that collisions, most of which are
grazing, produce only little gas but may efficiently replenish the
loss-cone for tidal disruptions.

In the past decade, gas-dynamical processes have been increasingly
favored over stellar dynamics as the main source of fueling of AGN
\citep[ and references therein]{SBF90,Shlosman92,Combes01}. It is
argued that, to achieve the highest QSO luminosities, the initial
stellar cluster has to be so dense that its formation is problematic
and would, most likely, require to concentrate a large amount of gas
in the galactic center anyway. Furthermore, whether most of the gas
emitted by stars --either in the course of their normal evolution or
through collisions-- finds its way to the MBH is uncertain (see
Sec.~\ref{subsec:futurework}).  However, it may have been overlooked
that the effective stellar relaxation rate, and, hence BH fueling
through tidal disruptions or direct horizon crossings, may be highly
enhanced by small departures from the assumption of a smooth spherical
potential. Such departures may be the presence of orbiting cores or
nuclear BHs of smaller accreted galaxies \citep{PR94,ZHR02}, or
triaxiality \citep{NS83} which may survive in the vicinity of the BH
even if it is destroyed at intermediate scales
\citep{PM01}.\footnote{Unfortunately, such possibilities, although
pointing to the importance of stellar dynamical processes, could only
be introduced approximately in our code which relies on spherical
symmetry.}

Even though purely stellar dynamical processes are probably only
secondary in feeding QSO-class MBHs, they may be efficient enough to
grow few million solar masses objects from BHs with a mass of a few
hundreds $M_\odot$. Furthermore, questions regarding the interplay
between the stellar nucleus and a central MBH are more pressing than
ever, as observational evidences for the presence of MBHs in most, if
not all, bright galaxies, including the Milky Way, are accumulating at
an impressive rate
\citep[][ and references in paper~I]{FPPMWJ01,MF00,KG01,GPEGO00,GMBTK00}. 
In particular, tidal disruptions at a rate of order
$10^{-4}\,\mathrm{yr}^{-1}$ seem unavoidable for BHs less massive than
a few $10^8\,M_\odot$, with the likely consequence of bringing back to
active life an otherwise quiescent galactic nucleus
\citep{Hills75,LO79,Rees88,Phinney89,SW93,SU99,MT99}. Ironically, while tidal 
disruptions are deemed too rare to be the main contributor to the
growth of MBHs, even in very dense nuclei, they are predicted in
present-day normal nuclei with a rate which is embarrassingly high in
regard to the low luminosity of these objects, a fact that has been
used to impose constraints on gas accretion models
\citep{SvO84,MQ01}. Some flaring events in the UV or X-ray band from
the center of active and non-active galaxies have been tentatively
interpreted as the accretional aftermath of tidal disruptions
\citep[][ and references therein]{GSZO00,Komossa01,Renzini01}. But 
further conclusions have to await more complete stellar dynamical
simulations, like the ones we propose to carry out with our code, and
a better understanding of the post-disruption accretion process in
order to predict its observational signature (wavelength, intensity,
duration, etc.)
Beside the accretion flares, another promising observational
consequence is predicted: the production of hot, very bright, stellar
cores of tidally stripped giant stars \citep{DSGMG01}.

As recalled above, collisions were first thought to play a major role
in very dense nuclei models, either by feeding the MBH
or by directly producing the AGN luminosity. The latter
class of models, now incorporating a central BH, has been revived by
\citet{TCFCP00} but should be examined in the light of a more
refined treatment of stellar collisions and stellar dynamics. Even if
they are not frequent enough to have a strong impact on the dynamics
or BH fueling, collisions may have interesting observational
consequences, by producing peculiar stellar populations, like blue
stragglers
\citep[][ and references therein, in the context of globular clusters]{SFLRW00}, 
or destroying giant stars \citep{GTKKTG96,Alexander99,BD99}, for
instance.

In addition to the now almost 'classical' questions concerning tidal
disruptions and collisions, the stellar dynamics of galactic nuclei is
key in other processes of high observational importance. An important
example is capture of compact stars on relativistic orbits around the
MBH. Through relaxation or collisions, a compact star may get on a
very elongated orbit with such a small pericenter distance that
emission of gravitational waves will drive further orbital evolution
until the star plunges through the horizon of the MBH
\citep{HB95,SR97,Freitag01,Ivanov01}. As these waves, if successfully detected
and analyzed, would be a direct probe to the space-time geometry near
MBHs \citep{Thorne98,Hughes01,Hughes01b}, such relativistic MBH-star
binaries will be prime-interest sources for the future space-borne
laser interferometer {\em LISA}
\citep{Danzmann00}.  This question and other ones to be mentioned 
in Sec.~\ref{subsec:futurework} are beyond the scope of this paper and the
relevant physics are not included in the code described here
\citep[see, however,][ for our first results concerning the capture
 of compact objects]{Freitag01}. Nonetheless, they strongly motivate
 the need for detailed numerical models of the stellar dynamics in the
 center-most parts of galaxies.

\subsection{General approach and limitations}

As is clear from this introduction and was already stressed in
Paper~I, the physics of galactic nuclei is a very intricate problem,
with dozen of physical processes or aspects that can potentially play
a role and interfere with each other. Any really general and realistic
approach would have to face too many computational challenges and
unknowns concerning the physics, initial and limit conditions to be
feasible at the present date. Various numerical methods have different
limitations and require different simplifying assumptions which
delineate the class of models that can be treated.

For instance, it is increasingly recognized that galaxy merging is a
common process in the universe and that such events have deep imprint
on the structure of galactic nuclei \citep{NM99,MC01}. Of particular
interest is the formation and evolution of binary BHs formed in the
process \citep{BBR80,GR00,HSS01,MM01,Yu01}. Self-consistent simulation
of these highly dynamical episodes in the life of galactic nuclei can
only be done with $N$-body codes in which the orbits of $N$ particles
are explicitly integrated for many dynamical times. However, such
direct $N$-body integrations are extraordinarily CPU-demanding and,
when various physical processes interplay whose relative importance
depends on $N$, their results can not be safely scaled to $N\gg 10^6$
to represent a real nucleus. Hence, even with cutting-edge special
purpose computers like GRAPE-6 \citep{Makino00}, $N$-body simulations
can not follow the evolution of a galactic nucleus over a Hubble time
if relaxation is appreciable.

The $N$ barrier can only be broken through by trading realism for
efficiency. This is done mainly through three core assumptions: {\bf
(1) Restricted geometry}: we assume that the nucleus is of perfect
spherical symmetry. {\bf (2) Dynamical equilibrium}: at any given
time, the system is a solution to the collisionless Boltzmann equation
\citep{BT87}.  {\bf (3) Diffusive 2-body relaxation}: the departures
from a smooth gravitational potential which is stationary on dynamical
time scales, are treated as a large number of uncorrelated 2-body
hyperbolic encounters leading to very small deflection angles. This is
the base of the standard Chandrasekhar theory of relaxation
\citep{Chandrasekhar60}.

To our knowledge, assumptions (2) and (3), which underlie the
derivation of the Fokker-Planck equation from the Boltzmann equation
\citep{BT87}, are shared by all methods aimed at simulating the
relaxational evolution of stellar clusters and all of them also rely
on spherical symmetry, with the exception of the code developed by
\citet{ES99} and \citet{KimEtAl01} which allows overall cluster rotation 
(see paper~I for a short review of these various methods). We have
based our code on the Monte Carlo (MC) scheme invented by H\'enon
\citeyearpar{Henon71a,Henon71b,Henon73,Henon75}. 
The reason for this choice, presented in detail in paper~I, is
basically that this algorithm offers the best balance between
computational efficiency, with CPU time scaling like
$N_\mathrm{p}\ln(cN_\mathrm{p})$ where $N_\mathrm{p}$ is the number of
particles and $c$ some constant, and the ease and realism with which
physics beyond relaxation, in particular stellar collisions, can be
incorporated. Other codes stemming from H\'enon's scheme have been
developed and very successfully adapted to the dynamics of globular
clusters
\citep{Stodol82,Stodol86,Giersz98,Giersz01,JRPZ00,WJR00,JNR01,FJPZR01} 
but we are not aware of any previously published adaptation of this
method to the realm of galactic nuclei.

\modif{
In principle, there is no other restriction on the initial conditions
for the cluster than conditions (1) and (2). In practice, however, the
code we use to build the initial cluster (see Appendix) is limited to
systems for which the distribution function (DF) depends on the energy
only and doesn't account for the presence of a BH at the center. The
first restriction implies that we cannot consider systems that present
initial velocity anisotropy or mass segregation. The second forces us
to start with 'seed' central BHs, i.e. the BH has to be initially so
light that its addition at the center of the nucleus doesn't
noticeably perturb the dynamical equilibrium. These limitations
correspond to the class of models that have been investigated in most
previous studies. For instance, with the exception of \citet{Rauch99},
all the self-consistent simulations cited above considered a seed BH
which grows through accretion of stellar matter. Even though this is
not a favored BH growth scenario anymore, in this paper, we adopt
such models mainly as a mean to establish the correct working of our
code through comparisons with the literature. 

If the central BH forms on a time scale much shorter than relaxation
time but longer than dynamical time, presumably through infall of gas
from outside the nucleus, as proposed by, e.g., \citet{vdM99c} and
\citet{McMH02}, the stellar cluster reacts {\em adiabatically}, a
process our code can cope with, as demonstrated in
Sec.~\ref{subsec:adiab}. This allows to create models at dynamical
equilibrium which contains a central BH of significant mass. Our
procedure for creating initial conditions can be adapted to clusters
with central BH for which the energy-dependent DF is known, such as
$\gamma$-models \citep{Tremaine94}. In recent simulations to be reported
in further papers, we use these models to investigate the dynamics of
present-day galactic nuclei. The aim of this approach is to gain
information about the rate and characteristics of interesting events
(collisions, tidal disruptions, captures\ldots) in $z\simeq 0$
galaxies without trying to guess which are the high-$z$ ``initial''
conditions. However, it is observationally, as well as theoretically,
doubtful that nearby galactic nuclei are devoid of anisotropy or mass
segregation\footnote{For instance, we find for a model of the central
cluster of the Milky Way, that significant segregation of stellar BHs
appears in less than 1\,Gyr so that assuming that no segregation has
occurred in the past history of the system is unrealistic.}. However,
the lack of published generalizations of the $\eta$-models, including
mass segregation and/or anisotropy, makes it difficult to test the
implications of these implicit assumptions.

The evolution of galactic nuclei is thought to go through highly
dynamical phases, most noticeably mergers with other nuclei predicted
by popular hierarchical structure formation scenarios. It is often
assumed that the central BHs formed as intermediate mass objects
($100-1000\,M_{\odot}$) and grew mainly by luminous gas accretion
during these episodes \citep[][ and references therein]{KaHa00,VHM02}
but the opposite view, i.e. that MBHs formed at high redshifts in the
core of only a small fraction of proto-galaxies and grew mostly by
merging together, cannot be ruled out \citep{MHN01}. Anyway, although
the MC code cannot follow these dynamical phases, one can easily use
the outcome of $N$-body simulations of such processes as initial
conditions, as soon as dynamical equilibrium is reached and the system
is reasonably spherically symmetric. The explicit knowledge of the
distribution function is not required here because one can directly
turn each $N$-body particle into one or a few super-star(s).}

\subsection{Units and definitions}
\label{sec:units}

When we do not explicitly indicate astrophysical
units, we use the ``code'' units defined in Sec.~3.2 of paper~I. 

$G$ is the gravitational constant. $M_\mathrm{BH}$ is the mass of the
central BH, $M_\mathrm{cl}$ the total stellar mass and $R_\mathrm{cl}$
the radius of the cluster (if finite). We use the following definition
for the core radius: $R_{\mathrm{c}}=\sqrt{9\sigma_0^2/4\pi G\rho_0}$
where $\sigma_0$ is the central 1D velocity dispersion and $\rho_0$ is
the central density of the cluster.

We assume the following relation for the Coulomb logarithm: $\ln
(\gamma N_\star)$ with $\gamma=0.4$ for single-mass models and
$\gamma=0.01$ when there is an extended stellar mass
spectrum. $N_\star$ is the total number of stars in the cluster. In
principle, the Coulomb logarithm, $\Lambda$ should be proportional to
$N_\star$ only if the cluster is self-gravitating. In a central region of radius
$GM_\mathrm{BH}\sigma_v^{-2}
\simeq R_\mathrm{cl}(M_\mathrm{BH}/M_\mathrm{cl})$ (assuming
$M_\mathrm{BH} \ll M_\mathrm{cl}$, $\sigma_v$ is the velocity
dispersion of the stars far from the BH), the BH gravitationally
out-weights the stellar cluster. There, the velocity dispersion at
distance $R$ of the center is $\sigma_v^2(R) \simeq GM_\mathrm{BH}/R$
and a steep cusp of stars is expected to develops so that,
$b_\mathrm{max} \simeq R$ is a sensible choice. Consequently,
according to Eq.~6 of paper~I, $\Lambda
\propto M_\mathrm{BH}/M_\star$ seems more
appropriate \citep{BW76,LS77,MEG00}. We have conducted test
calculations with a $R$-variable Coulomb ratio set to $\Lambda \propto
T_{\mathrm{orb}}(R)/T_{\mathrm{min}}(R)$ where $T_{\mathrm{orb}}
\approx (GM_r/R^3)^{-1/2}$ is a measure of the orbital time and
$T_{\mathrm{min}}$ corresponds to the shortest effective 2-body
encounter, i.e. $T_{\mathrm{min}} \approx b_0/\sigma_v \approx GM_\ast
\sigma_v^{-3}$. Such a choice is motivated by the fact that a
transient potential fluctuation with time scale much longer than
$T_{\mathrm{orb}}$ will act adiabatically on the motion of a given star
and thus leave its orbit unchanged after it is over. Results are not
significantly affected by the choice of $\Lambda$, which convinced us
to keep the simple $\Lambda =\gamma N_{\ast}$ relation.

\section{Stellar collisions}
\label{sec:coll}

\subsection{Use of SPH collision simulations}

The inclusion of realistic collisions\footnote{Here, by ``collision'',
we mean a genuine hydrodynamical contact encounter between two stars,
as opposed to mere 2-body gravitational deflections.} is probably the
main improvement over previous cluster evolution codes that our scheme
features. In the past few years, we have been computing thousands of
3D hydrodynamics simulations of collisions between MS stars
using a SPH code
\citep{Benz90}. For simplicity, only collisions between main sequence 
stars are considered for the time being. Actually, giant stars are
expected to dominate the collision rate \citep{BD99,FB02}. The effects
of collisions are included in the cluster simulations with unpreceded
realism by interpolating the outcome of these events from the huge
SPH-generated results database
\citep{FB02}.  In so doing, we get rid of many of the uncertainties
introduced by the simplistic recipes formerly used in simulations of
collisional cluster dynamics.  Unfortunately, even with such a
procedure, important simplifications of the physics have still to be
done.  The major ones are connected with the possible formation of
binaries through tidal dissipation of orbital energy and to the
stellar evolution of the stars that have suffered from a collision. We
discuss both problems in turn.

The cross sections for the formation of so-called ``tidal-binaries''
are not well known \citep{PT77,LO86,MMMDT87,BH92,LRS93,KL99} and their
long-term evolution is still debated \citep{BH92,LRS93}. Hence, it is
fortunate that the rate of tidal captures is overtaken by the rate of
collisions as soon as $\sigma_v/v_{\ast}>0.1$ where $\sigma_v$ is the
1D velocity dispersion of the stars and
$v_{\ast}=\sqrt{2GM_{\ast}/R_{\ast}}$ is the escape velocity at the
surface of a star \citep[ Fig.~16]{LRS93}. Consequently, as we expect
quite high stellar velocities in the center of galactic nuclei
(particularly near a super-massive BH), we decided to neglect tidal
capture in our code.

A parameter of prime importance is the star's radius as it determines
its collisional cross section and, hence, the probability of
subsequent collisions that could lead, for instance, to the runaway
build-up of more and more massive stars by multiple mergers. After a
collision, as a large amount of energy has been injected into the
stellar envelope, the star is much larger than a MS star with the same
mass. However, on a Kelvin-Helmholtz time scale ($T_{\mathrm{KH}}$)
the radius shrinks back to the MS value, as the stellar structure
returns to thermal equilibrium. Here we assume $T_{\mathrm{coll}} \gg
T_{\mathrm{KH}}$ so we can neglect the short swollen phase and attribute a
MS radius to the collision product.

When stellar evolution is taken into account, it becomes in principle
necessary to know what amount of collisional mixing occurs and how it
affects the MS life-time $T_{\mathrm{MS}}$ of the product. We can
expect that, contrary to parabolic mergers where only little mixing
takes place
\citep[ for instance]{SLBDRS97}, high velocity collisions are able to
rejuvenate the star by bringing fresh hydrogen-rich gas from the outer
parts to the center. If two stars of unequal masses merge together,
simulations show that the smaller one, whose material is of lower
entropy, sinks to the center of the larger one \citep{LWRSW01}. This
appears as an efficient mechanism to bring fuel directly to the core
of the large star and delay hydrogen exhaustion.  Conversely, the
higher mean molecular weight $\mu$ that results from spreading the
central Helium (produced by H-burning on the MS) leads to a important
decrease of $T_{\mathrm{MS}}$ as compared to a star with a ``normal''
composition
\citep{CG98}.  Indeed, from homological relations, one finds:
$T_{\mathrm{MS}} \propto \mu^{-4}$ \citep{KW94}. On the other hand,
the radius depends only weakly on $\mu$ ($R \propto \mu^{0.6}$ for the
CNO-cycle) so we can safely neglect the effects on the collision cross
section in our simulations\footnote{How the outcome of further
  collisions will be influenced by structural changes due to previous
  collisions has not yet been assessed. This can be of importance in
  the case of ``run-away'' mergers.}.

\subsection{Collision rate.}
\label{sec:coll_rate}

Let's consider a close approach between two stars with masses and
radii $M_1$, $R_1$ and $M_2$, $R_2$, respectively. The relative
velocity at infinity is $v_{\mathrm{rel}}$ and the impact parameter
$b$. A collision occurs when the centers of the stars are closer to
each other than $d = \eta(R_1+R_2)$ ($\eta=1$ for genuine collision,
$\eta\leq1$ for merging, $\eta\geq1$ for tidal capture when
$v_{\mathrm{rel}}$ is small enough). Until this collision distance is
reached, we neglect the gravitational influence of other stars as well
as any mutual tidal interaction. So the problem reduces to a simple
hyperbolic approach between two point masses. This gives us, the
largest impact parameter leading to contact, $b_\mathrm{max}$, and the
cross section,
\begin{equation}
  S_{\mathrm{coll}}^{(12)} = \pi b_{\mathrm{max}}^2 = 
  \pi \eta^2(R_1+R_2)^2 \left( 1 +
      \frac{\left(v^{(12)}_{\ast}\right)^2}{\eta v_{\mathrm{rel}}^2} 
  \right)
  \label{eq:Scoll}
\end{equation}
where 
\begin{equation}
  v^{(12)}_{\ast}=\sqrt{\frac{2G(M_1+M_2)}{R_1+R_2}}.
  \label{eq:Vesc_stell}
\end{equation}
The second term is the bracket of equation~\ref{eq:Scoll} is the
gravitational focusing which enhances the cross-section at low
velocity ($S^{(12)}_{\mathrm{coll}} \propto R_1+R_2$). At high
velocities $S_{\mathrm{coll}}$ tends to the geometrical value $\pi
f^2(R_1+R_2)^2$. So, the collision rate for a test-star ``1'' in a
field of stars ``2'' having all the same mass, radius and relative
velocity to ``1'' is simply
\begin{equation}
  \left. \frac{\mathrm{d}N_{\mathrm{coll}}}{\mathrm{d}t}\right|^{(1;2)} = 
  n_2 v_{\mathrm{rel}} S_{\mathrm{coll}}^{(12)}
  \label{eq:CollRate1}
\end{equation}
where $n_2$ is the (local) number density of stars ``2''.

If we are interested in the overall collision rate in a star cluster,
the next step to do is to introduce a velocity distribution. Before
considering more general cases, let's assume that all stars in the
cluster have the same mass $M_{\ast}$ and radius $R_{\ast}$ (so we can
drop over-score ``$^{(12)}$'' in $v_{\ast}$ and $S_{\mathrm{coll}}$)
and that their density is $n_\ast$.  The average local collision time
$T_{\mathrm{coll}}(R)$ is found by integrating the collision rate
(equation~\ref{eq:CollRate1}) over all possible velocities of the two
stars,
\begin{equation}
  \frac{1}{T_{\mathrm{coll}}} = \frac{1}{n_\ast} \int 
  \mathrm{d}^3\vec{v}_1 \mathrm{d}^3\vec{v}_2 
  f(\vec{v_1}) f(\vec{v}_2) \|\vec{v}_1-\vec{v}_2\| 
  S_{\mathrm{coll}}.
  \label{eq:IntTcoll1}
\end{equation}
As shown in \citet{BT87}, the result for a
Maxwellian distribution is
\begin{equation}
  \left. \frac{1}{T_{\mathrm{coll}}} \right|_{\mathrm{M}} = 
  \underbrace{16\sqrt{\pi}}_{\displaystyle \simeq 28.4} \eta^2 R_{\ast}^2 n_\ast \sigma_v 
  \left(1+ \frac{v_{\ast}^2}{4\eta \sigma_v^2} \right).
  \label{eq:TcollM}
\end{equation}
For the Plummer model, the result is very similar to
Eq.~\ref{eq:TcollM}, with only the numerical constant replaced by
$28.6$.

\begin{figure}
  \resizebox{\hsize}{!}{
    \includegraphics{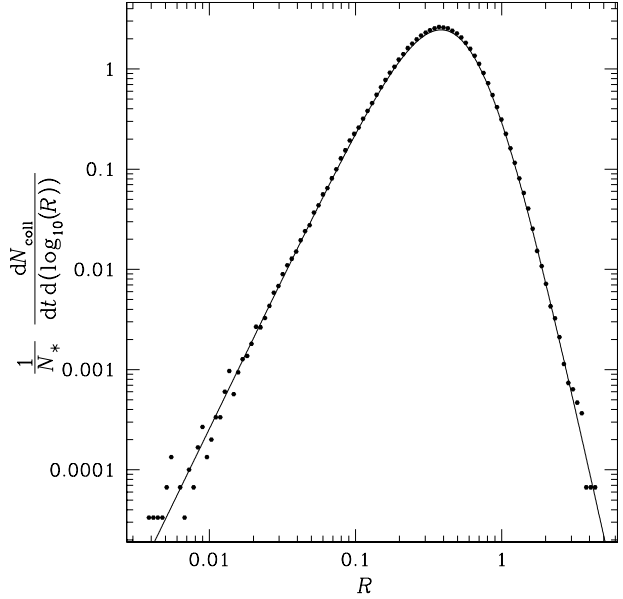}
    }
  \caption{
    Collision rate as a function of radius in a Plummer cluster with
    $\Theta_0 = 0.725$ and $N_\ast=10^6$. The solid line is the
    theoretical rate based on Eq.~\ref{eq:dNcolldRPlum}. The dots are
    statistics from a MC simulation run with no cluster evolution.
    ``N-body units''
    are used (see Sec.~\ref{sec:units}). }
  \label{fig:dNcolldRPlum}
\end{figure}

The total number of collisions per unit time in the cluster is given
by the integration of $1/T_{\mathrm{coll}}(R)$ over the whole cluster:
\begin{equation}
  \left. \frac{\mathrm{d}N_{\mathrm{coll}}}{\mathrm{d}t}\right|_{\mathrm{tot}} =
  \frac{1}{2}\frac{N_{\ast}}{\widehat{T}_{\mathrm{coll}}} =
  2 \pi \int_0^{\infty} \mathrm{d}R\,R^2 n_\ast(R)
  \frac{1}{T_{\mathrm{coll}}(R)}.
  \label{eq:IntdNcolldt}
\end{equation}
For a Plummer model of total mass $M$, star number $N_{\ast}$ and
scale radius $R_\mathrm{P}$, the collision rate by
unit radius reads:
\begin{eqnarray}
  \frac{ \mathrm{d}N_\mathrm{coll} }{ \mathrm{d}t\,\mathrm{d}R }(R) &=&
  54\sqrt{2}\frac{ \sqrt{G \rho_0} }{ R_\mathrm{P} }
  \frac{1}{\Theta_0^2} \times \nonumber \\
  && 
  u^2 \left(1+u^2\right)^{-21/4} \left[ 1+
  \Theta_0 \left(1+u^2\right)^{1/2} \right]
\label{eq:dNcolldRPlum}
\end{eqnarray}
with 
\[ 
  u = \frac{R}{R_\mathrm{P}}, \;\;
  \rho_0 = \rho(0) = \frac{3}{4\pi}\frac{M}{R_\mathrm{P}^3},
\]
and
\[
  \Theta_0 = \frac{v_{\ast}^2}{4\eta \sigma_v^2(0)} = 
  \frac{3}{\eta N_{\ast}} \frac{R_\mathrm{P}}{R_{\ast}}
  \mbox{~~(Safronov number)}.
\]
As a check of our code, Fig.~\ref{fig:dNcolldRPlum} depicts this rate
along with the statistics produced in a inventory run during which
the cluster's structure as a Plummer model was frozen.

Carrying out the radial integration, we finally get the total
collision rate in the whole cluster:
\begin{equation}
  \left. \frac{\mathrm{d}N_{\mathrm{coll}}}{\mathrm{d}t}\right|_{\mathrm{tot}} =
  \sqrt{G\rho_0}\frac{1}{\Theta_0^2}\left(4.25+5.20\Theta_0\right)
\end{equation}.


\subsection{Relative collision rates between stars of different masses.}
\label{sec:rel_coll_rate_m1m2}

\begin{figure*}
  \resizebox{12cm}{!}{
    \includegraphics{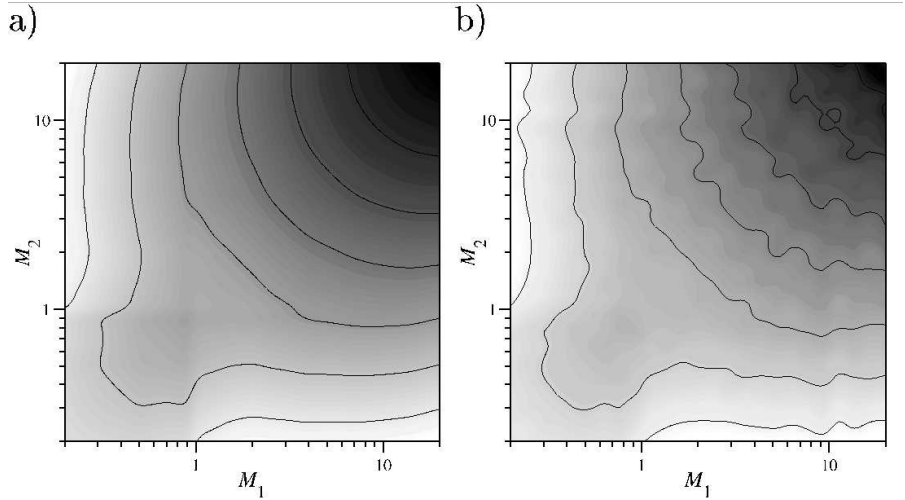}
    }
  \hfill
  \parbox[b]{55mm}{
    \caption{ 
      Total relative collision rate $\Gamma_\mathrm{tot}(M_1,M_2)$
      between stars with masses $M_1$ and $M_2$ in a Plummer cluster
      without mass-segregation. The gravitational focusing parameter
      is $\widehat{\Theta} = 1.5$. Masses are in $M_{\sun}$.
      Lighter gray shades correspond to higher values. Successive
      contour levels correspond to factor of 2 decrease in $\Gamma$.
      The same levels are drawn on both panels. The mass-function is
      $\Psi(M_\ast) \propto M_\ast^{-1.35}$ for $M_\ast$ between $0.2$
      and $20 M_{\sun}$ and the Mass-Radius relation is set according
      to stellar models by \protect\citet{SSMM92} and
      \protect\citet{CDSBMMM99} {\bf a)}
      Theoretical rate from Eq.~\protect\ref{eq:TotRelCollRateMiMj}.
      {\bf b)} Statistics from the MC code (400\,000
      collisions) with cluster evolution inhibited. This comparison
      demonstrates the accuracy of the collision sampling in our
      code.}  
  \label{fig:TotRelCollRateMiMj}
    }
\end{figure*}

We now address the case of a cluster with a distribution of stellar
masses. For simplicity, we consider a discrete mass spectrum with
$N_{\mathrm{sp}}$ components: $M_{\ast} \in \{ M_i
\}_{i=1}^{N_{\mathrm{sp}}}$ with (local) densities $n_i$. So, using
Eq.~\ref{eq:CollRate1}, the rate by unit volume for collisions
between stars of classes $i$ and $j$, with velocities $\vec{v}_i$ and
$\vec{v}_j$ is 
\begin{equation}
  \label{eq:CollRateij1}
  \Gamma_{ij}\mathrm{d}^3\vec{v}_i\mathrm{d}^3\vec{v}_j =
  f_i(\vec{v}_i)f_j(\vec{v}_j)\|\vec{v}_i-\vec{v}_j\|S_{\mathrm{coll}}^{(ij)}
  \mathrm{d}^3\vec{v}_i\mathrm{d}^3\vec{v}_j
\end{equation}
where $f_i$, $f_j$ are the phase-space DFs which are
assumed to comply with (spatial) spherical symmetry and isotropy.
Their $R$-dependence is implicit. If we further assume Maxwellian
velocity distributions with 1-D velocity dispersions $\sigma_i$ and
$\sigma_j$, the distribution of the relative velocity
$\vec{v}_{\mathrm{rel}}=\vec{v}_i-\vec{v}_j$ is Maxwellian too, with
dispersion $\sigma_{ij}=\sqrt{\sigma_i^2+\sigma_j^2}$. We keep
$v_{\mathrm{rel}}=\|\vec{v}_{\mathrm{rel}}\|$ as the only relevant
velocity variable by integrating Eq.~\ref{eq:CollRateij1} over the
others:
\begin{eqnarray}
  \label{eq:CollRateij2}
  \lefteqn{ \Gamma_{ij}\mathrm{d}v_{\mathrm{rel}} \propto 
  n_i n_j \left(R_i+R_j\right)^2 v_{\mathrm{rel}}^3
  \left( 1 + \left(\frac{v^{(ij)}_{\ast}}{v_{\mathrm{rel}}}\right)^2
  \right) \times } \nonumber \\
  & &
  \sigma_{ij}^{-3}
  \mathrm{e}^{-\frac{1}{2}\frac{v_{\mathrm{rel}}^2}{\sigma_{ij}^2}}
  \mathrm{d}v_{\mathrm{rel}}. 
\end{eqnarray}
$\eta$ has been set to 1.  For a continuous mass spectrum, we define
the mass function as
$\psi(M_i) = n_\ast^{-1} \mathrm{d}n_\ast(M_i)/\mathrm{d}(\log_{10}(M_i))$ so we
have to substitute $\mathrm{d}n_i =
n_\ast \psi(M_i)\mathrm{d}(\log_{10}(M_i))$ for $n_i$ in the previous
formula. In order to get a equation for the relative collision rate
between stars of different masses (per unit volume, $\log_{10}(M_i)$ and
$\log_{10}(M_j)$), we assume $\sigma_i=\sigma_j=\sigma_v \forall i,j$
and integrate over $v_{\mathrm{rel}}$:
\begin{equation}
  \label{eq:CollRateMiMj3}
  \Gamma(M_i,M_j) \propto
  n_\ast^2 \sigma_v \psi(M_i)\psi(M_j) \left(R_i+R_j\right)^2
  \left( 1 + \Theta^{(ij)} \right)
\end{equation}
\[
\mbox{with~~} \Theta^{(ij)} =
  \frac{\left(v^{(ij)}_{\ast}\right)^2}{4\sigma_v^2} =
  \frac{G\left(M_i+M_j\right)}{2\sigma_v^2\left(R_i+R_j\right)}.
\]
For a Plummer model with no mass-segregation (and, thus, a unique
$\sigma_v(R)$), this relation, when integrated over the whole cluster,
leads to
\begin{eqnarray}
  \Gamma_\mathrm{tot}(M_i,M_j) & = &
  \frac{ \mathrm{d}N_\mathrm{coll} }{
    \mathrm{d}t \, \mathrm{d}(\log_{10}{M_i}) \, \mathrm{d}(\log_{10}{M_j}) } 
   \nonumber \\
 & \propto & \psi(M_i)\psi(M_j) \sqrt{G\rho_0} N_\ast^2 \left( \frac{ R_i+R_j 
    }{ R_\mathrm{P} } \right)^2 \nonumber \\
 \label{eq:TotCollRateMiMj}
 & & \mbox{} \times \left( 1 + 3.66 \frac{M_i+M_j}{M}
  \frac{R_\mathrm{P}}{R_i+R_j} \right) \\
 & \propto & \psi(M_i)\psi(M_j) \left(\frac{ R_i+R_j }{ R_\odot }\right)^2
 \nonumber \\
 \label{eq:TotRelCollRateMiMj}
 & & \mbox{} \times \left( 1 + \widehat{\Theta} \frac{
     (M_i+M_j)/M_{\sun} }{ (R_i+R_j)/R_{\sun} } \right)
\end{eqnarray}
\[
 \mbox{with~~} \widehat{\Theta} = 3.66 \frac{R_\mathrm{P}/R_{\sun} }{
   M/M_{\sun} }.
\]
In relation \ref{eq:TotRelCollRateMiMj}, only the dependencies on
stellar quantities have been preserved to insist on the relative
collision rates between different stellar species.  Although it relies
on the admittedly unrealistic hypothesis of no mass segregation,
Eq.~\ref{eq:TotRelCollRateMiMj} proves useful as a prediction our code
can easily (and successfully) be tested against, see
Fig.~\ref{fig:TotRelCollRateMiMj}.

\subsection{Introduction of stellar collisions in the MC code}
\label{sec:MC_coll_sim}

%
%
%
%
The difficulty of introducing stellar collisions in any stellar
dynamics code is twofold. First, as the previous discussion has shown,
it is not at all straightforward to determine the correct distribution
of collision parameters ($v_{\mathrm{rel}}$, star types, position in
the cluster, \ldots). Secondly, provided the result of a particular
collision is known (by performing hydrodynamical simulations, for
instance), we want to be able to preserve as much as possible of that
valuable information when introducing it back in the cluster evolution
code. Due to their very structure\footnote{Their basic limitation lies
in the principle they owe their efficiency to: they model the
stellar system as a set of continuous DFs (one for each different
stellar mass).}, some widely used schemes, based on an explicit
resolution of the Fokker-Planck equation, impose such a highly
simplified treatment of the collisions' outcome that it would not make
much sense to devote energy to a realistic computation of these
events. The MC method is exempt of such limitations.

\subsubsection{Global code modifications.}

Collisions introduce a new time scale in the code. There is
consequently a new constraint on the time steps 
\begin{equation}
  \delta t(R) \leq f_{\delta t} \tilde{T}_\mathrm{coll}(R)
  \label{eq:ineq_dt_coll}.
\end{equation}
$\tilde{T}_\mathrm{coll}(R)$ is an estimation of the local collision
time. We chose the following definition, based on Eq.~\ref{eq:TcollM}:
\begin{equation}
  \frac{1}{\tilde{T}_\mathrm{coll}(R)} = 
  16\sqrt{\pi} n_\ast \sigma_v \langle R_\ast^2 \rangle
  \left[ 1 + \frac{G \langle M_\ast R_\ast \rangle}{
      2 \sigma_v^2 \langle R_\ast^2 \rangle} \right]
\end{equation}
where $\sigma_v^2=\langle v^2 \rangle$ and
$\langle\mbox{bracketed}\rangle$ quantities are local averages. This
particular expression was chosen for its ease of evaluation and
because, provided all stellar species have isothermal velocity
distribution (quite a strong demand!), it reduces to exact relations
in the two interesting limiting cases:
\begin{equation}
  \tilde{T}_\mathrm{coll}^{-1} \rightarrow \langle T_\mathrm{coll}^{-1} \rangle
  = \left\{
      \begin{array}{lrl}
          {\displaystyle 8\sqrt{\pi} G n_\ast \frac{\langle M_\ast R_\ast
              \rangle}{\sigma_v} }
          &\mbox{for}&
          \sigma_v^2 \ll \langle v_\ast^2 \rangle \\[10pt]
          {\displaystyle 16\sqrt{\pi} n_\ast \sigma_v \langle R_\ast^2
            \rangle }
          &\mbox{for}&
          \sigma_v^2 \gg \langle v_\ast^2 \rangle.
      \end{array}
    \right.
\end{equation}
By requiring
\begin{equation}
  \delta t(R) \leq f_{\delta t} \left( \tilde{T}_\mathrm{rel}^{-1}(R) +
    \tilde{T}_\mathrm{coll}^{-1}(R) \right)^{-1},
\end{equation}
we make sure that time steps are short enough to resolve both
relaxational and collisional processes. Apart for this extended
constraint, all the time-step determination and pair selecting
machinery of paper~I is left formally unchanged.

\subsubsection{Monte Carlo sampling of the collisions.}

Relaxation is due to the cumulative effects of a huge number a small
individual scatterings and can be treated as a continuous process,
affecting progressively the particles' orbits. To be computationally
tractable this phenomenon is discretized back into
``super-encounters''. In contrast, collisions do not act gradually but
are genuinely discrete events, each of which strongly affect the
properties of the implied stars.  Hence, there seems to be no way to
add up the effects of collisions into ``super-collisions'', no escape
from the necessity to simulate them as individual events.

When a pair of adjacent super-stars is selected to be evolved for a
time step $\delta t$, we randomly orient their velocities and compute
the local number density of stars of any kind, $n_\ast$, as explained
in paper~I. The probability for a mutual collision to occur during
that time span is, adapting Eq.~\ref{eq:CollRate1},
\begin{equation}
  P_\mathrm{coll}^{(12)} = N_\mathrm{coll}^{(12)} = n_\ast
  v_{\mathrm{rel}} S_{\mathrm{coll}}^{(12)}\delta t.
  \label{eq:CollProb}
\end{equation}
When compared to Eq.~\ref{eq:CollRate1}, this expression could be
thought to be an overestimate as $n_\ast$ is used instead of $n_2$.
Actually, for a given super-star of type ``1'', the expectation value
for the number of collisions with super-stars of type ``2'' is
\begin{eqnarray}
  \left\langle N_\mathrm{coll}^{(12)} \right\rangle & = &
  \underbrace { \left[\parbox{2.4cm}{Prob. for neighbor of being of
        type 2}\right] }_{\displaystyle n_2/n_\ast} \cdot
  \underbrace { \left[\parbox{2.4cm}{Collision prob. if neighbor is of type
      2}\right] }_{\displaystyle P_\mathrm{coll}^{(12)}} \nonumber \\
  & = & n_2 v_{\mathrm{rel}} S_{\mathrm{coll}}^{(12)} \delta t,
\end{eqnarray}
as needed. 
%
%
The collision probability is compared with a random number
$X_{\mathrm{rand}}$ with $[0;1[$-uniform deviate. If
$X_{\mathrm{rand}} <P_\mathrm{coll}^{(12)}$, a collision has to be
simulated whose initial conditions are completely determined as soon
as a value of the impact parameter $b$ has been chosen according to
probability density
\begin{equation}
  \mathrm{d}P = \left\{
      \begin{array}{ll}
        {\displaystyle \frac{2b\,{\mathrm{d}b}}{b^2_\mathrm{max}} } & 
        \mbox{if $0 \le b \le b_\mathrm{max}$,} \\
        0 &  \mbox{otherwise.}
      \end{array}
      \right.
\end{equation}
Note that the super-star pair is tested for collisions before
relaxation is applied to it. In case a collision is suffered, the
orbits are probably deeply modified. So the relaxation step is
 skipped even if the pair survived.

\subsubsection{Treatment of an individual collision.}

As explained earlier, the outcome of collisions happening in the
course of the cluster's evolution is specified by a large set of 3D
hydrodynamical simulations. These are potentially able to provide us
with any detail, significant or not, about the state of the resulting
star(s) and released gas. Most of this information, however is of no
real relevance so we focus on the important parameters we have to sort
out of this data and plug into the MC code. In another paper
\citep{FB02}, we describe the way collisions are simulated with an SPH
code and how we extract the needed ``macroscopic'' information back
from the simulation. Suffice to say that, if we assume the center of
mass (CM) reference frames defined \emph{before} and \emph{after} the
collision are the same (i.e. that $M^{\prime}_1\vec{w}^{\prime}_1 +
M^{\prime}_2\vec{w}^{\prime}_2 = 0$ where $M^{\prime}_{1,2}$ and
$\vec{w}^{\prime}_{1,2}$ are the post-collision masses and velocity
vectors in the pre-collision CM frame), the kinematic outcome is
entirely described by 4 numbers. They are $M_1^{\prime}$,
$M_2^{\prime}$, the final relative velocity at infinity,
\[
  v^{\prime}_{\mathrm{rel}} =
  \sqrt{\frac {2E^{\prime}_{\mathrm{orb}}}
    {M^{\prime}_1M^{\prime}_2/\left(M^{\prime}_1+M^{\prime}_2\right)} }
\]
and the deflection angle $\theta_{\mathrm{coll}}$. Further information
is contained in the post-collision stellar structure but it may be
ignored if one assumes, as we do, that the produced star(s) return to
normal MS structure. These 4 numbers are all we need to implement
collisions between super-stars following exactly the same scheme as
described in Sec.~4.2.1 of paper~I (steps 2--4) for purely
gravitational encounters. The only added difficulty is connected with
mass changes and the proper tracking of energy variation they
imply. 

\modif{Note that when a collision between two super-stars occurs, it amounts
to {\em each} star in the first super-star colliding with a star from
the second super-star. As the number of stars per super-star is the
same by construction, one can apply the outcome of the collision (new
mass and velocity) uniformly to all stars of the super-star,
i.e. to the super-star as a whole. When the stellar collision results
in to surviving stars, we have to modify the orbital and stellar
properties of both super-stars; when there is only one star left
(merger or destruction of the smaller star, see \citealt{FB00c}), one
superstar is removed and the other one is given the properties of the
remaining star; if both stars are destroyed, both super-stars are
removed from the simulation.}

\section{Tidal disruptions}
\label{sec:disr}

\subsection{Loss cone theory}
\label{sec:disr_lc}

If a star ventures very close to the BH, it may be broken
apart by tidal forces. The condition for an element of mass to be
stripped away from the surface of the star is that the instantaneous
gravitational attraction on it (due to the BH and the star itself) be
lower than the required centripetal acceleration. In the simplified
case of a non-rotating\footnote{In case of a co-rotating spherical
  star on a circular orbit, one gets a factor 3 instead of 2 inside
  $(\cdots)^{1/3}$.} spherical star on a Keplerian orbit,
this condition determines the following disruption radius:
\begin{equation}
 R_\mathrm{disr} \simeq \left(2\frac{M_\mathrm{BH}}{M_\ast} \right)^{1/3}
 \!\! R_\ast = \left(\frac{3}{2\pi} \frac{M_\mathrm{BH}}{\rho_\ast} \right)^{1/3}.
\label{eq:Rdisr}
\end{equation}
Where $\rho_\ast$ is the average density of the stellar matter. This
approximation assumes $M_\mathrm{BH} \gg M_\ast$. Note that this is
really only the condition for the tidal stripping of the outer layers
of gas because the stellar density increases toward the center of the
star. A more realistic approach should account for elliptical or
parabolic orbits, tidally induced deformation and the genuine
hydrodynamical nature of this violent phenomenon. Moreover, if deep
encounter certainly result in complete star destruction, milder ones
would be responsible of partial envelope stripping. Many studies have
addressed these aspects
\citep{CL83,EK89,LMZD93,Fulbright96,ALP00}. Fulbright performed SPH
simulations of parabolic encounters whose strength can be
parameterized by
\begin{equation}
  \beta = \frac{R_\ast}{R_\mathrm{peri}}
  \left(\frac{M_\mathrm{BH}}{M_\ast}\right)^{1/3}. 
  \label{eq:beta_tidal}
\end{equation}
For polytropic star models with $n=3/2$ and $n=3$, he found that
stripping of half the stellar mass occurs for $\beta_\mathrm{h} \simeq
0.8$ and $\beta_\mathrm{h} \simeq 1.7$, respectively. In the present
version of our code, complete disruption is assumed for $\beta >
\beta_\mathrm{h}$ while the star is left undamaged for more distant
encounters. This corresponds to Eq.~\ref{eq:Rdisr} with the factor
$2^{1/3}$ replaced by $\beta_{\mathrm{h}}^{-1}$.

\begin{figure}
  \resizebox{\hsize}{!}{
    \includegraphics{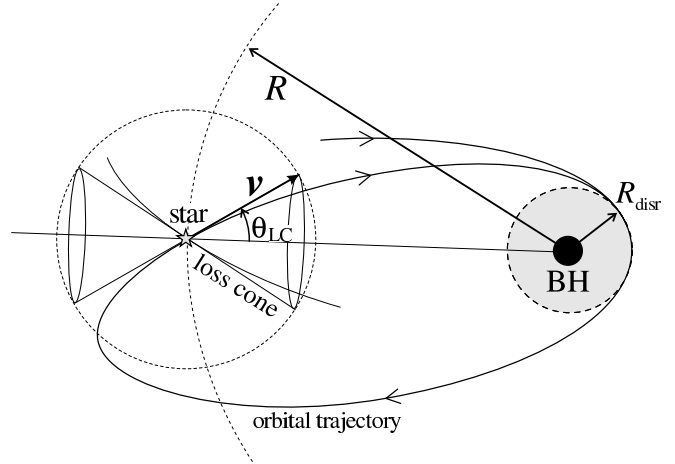}
    }
  \caption{ Diagram of the loss cone. }
  \label{fig:orbit_lc}
\end{figure}

The ``loss orbits'' are the set of stellar orbits with pericenter
distance $R_\mathrm{peri}$ smaller than $R_\mathrm{disr}$. For a star at
distance $R$ to the center with velocity modulus $v$, the \emph{loss
  cone} (LC) is the set of velocity directions that leads $R_\mathrm{a} <
R_\mathrm{disr}$, either going to the BH our coming from it (see
Fig.~\ref{fig:orbit_lc}). The aperture angle of the loss-cone,
$\theta_\mathrm{LC}$, is given by the relation
\begin{eqnarray}
  \label{eq:lc_angle_exact}
  \sin^2(\theta_\mathrm{LC}) &=& 2 \left( \frac{R_\mathrm{disr}}{vR}
  \right)^2 \left[\frac{v^2}{2} + \frac{GM_\mathrm{BH}}{R_\mathrm{disr}}
    \left( 1-\frac{R_\mathrm{disr}}{R} \right) \right. \nonumber \\*
  && \left. \rule{0pt}{14pt} \mbox{} + \Phi_\ast(R) -
    \Phi_\ast(R_\mathrm{disr}) \right]
\end{eqnarray} 
where $\Phi_\ast(R) = \Phi(R)+GM_\mathrm{BH}/R$ is the cluster
contribution to the gravitational potential. As, for reasonable
parameters, $R_\mathrm{disr}$ is a tiny value, typical loss orbits are 
very elongated, so that $R\gg R_\mathrm{disr}$ and
$GM_\mathrm{BH}/R_\mathrm{disr} \simeq v_\ast^2
(M_\mathrm{BH}/M_\ast)^{2/3} \gg v^2$. Hence Eq.~\ref{eq:lc_angle_exact}
simplifies to
\begin{equation}
  \label{eq:lc_angle_approx}
  \theta_\mathrm{LC}^2 \simeq 2\frac{ GM_\mathrm{BH}R_\mathrm{disr} }{ 
    v^2R^2 }.
\end{equation}
The loss cone is usually very small, as is demonstrated by an
order-of-magnitude estimate of $\theta_\mathrm{LC}$ at the BH's
``influence radius'' ($R_\mathrm{i}=GM_\mathrm{BH}/\sigma_v^2$):
\begin{eqnarray}
  \theta_\mathrm{LC}^2(R_\mathrm{i}) &\simeq& 
  N_\ast \left(\frac{ M_\ast }{ M_\mathrm{BH} }
  \right)^{2/3} \frac{R_\ast}{R_\mathrm{h}} \\*
  &\simeq& 2\times10^{-5} \left(\frac{N_\ast}{10^7}\right)
  \left(\frac{M_\mathrm{BH}}{10^6 M_{\sun}}\right)^{-\frac{2}{3}}
  \left(\frac{R_\mathrm{h}}{1 \mathrm{pc}}\right)^{-1} \\*
 && \mbox{for $R_\ast=1 R_{\sun}$ and $M_\ast=1 M_{\sun}$.} \nonumber
\end{eqnarray}
$R_\mathrm{h}$ is the cluster's half-mass radius.

If it wasn't for relaxation or other orbit modifying mechanisms
(collisions for instance) these loss orbits, if initially populated,
would be drained over a dynamical time and no further tidal disruption
would be expected in the subsequent cluster evolution, unless some
increase of $R_\mathrm{disr}$ occurs. This could happen for the whole
cluster as a result of the BH accreting gas supplied to it by other
sources like stellar winds ($M_\mathrm{BH} \nearrow$), or, as
investigated by \citet{SU99}, for those stars that
experience rapid swelling when they become red giants ($R_\ast
\nearrow$).

The crux of determining the rate of tidal disruptions, however, is the
role of relaxation. This process is capable of replenishing loss cone
orbits while at the same time it can remove stars from such orbits
thus preventing them from being disrupted. These effects have been
tackled either using quite rigorous approaches \citep{LS77,CK78,MT99}
mainly aimed at their inclusion into Fokker-Planck codes, or resorting
to more approximate descriptions \citep{FR76,SU99,MEG00}.  Here we
only outline the problem by recalling a few simple facts.

Eq.~\ref{eq:lc_angle_approx} can be recast in a simple
characterization of loss orbits:
\begin{equation}
  J^2 \le J_\mathrm{LC}^2 \simeq 2 GM_\mathrm{BH}R_\mathrm{disr},
  \label{eq:J2LC}
\end{equation}
a condition independent of energy $E$ (for stars not too tightly bound
to the BH). Thus the flux of stars to/from disruption orbits is
chiefly controlled by $J$-``diffusion'' in the vicinity of the
$J_\mathrm{LC}$ borderline. For a given star, let $\delta
J_\mathrm{orb}$ be the mean quadratic variation of the angular
momentum due to relaxation during a single orbit (defined as the
trajectory segment from a passage to apocenter position to the next
one),
\begin{equation}
  \delta J_\mathrm{orb} =\sqrt{ \left \langle (\Delta J)^2 
  \right \rangle_{P_\mathrm{orb}} }.
\end{equation}
If $\delta J_\mathrm{orb} \gg J_\mathrm{LC}$, stars can survive many
orbits, scattered into and out of loss trajectories before being
tidally disrupted. It follows that orbits with $J< J_\mathrm{LC}$ are
not strongly depleted and this regime is referred to as \emph{full loss
  cone}. If the velocity distribution is initially isotropic, this
process doesn't modify that fact and the fraction of stars disrupted
per orbital period is simply those of velocity directions pointing in
the loss-cone:
\begin{equation}
  \frac{\mathrm{d}\dot{N}_\mathrm{full}}{\mathrm{d}N} =
  \frac{1}{P_\mathrm{orb}}
  \frac{\theta_\mathrm{LC}^2}{2}.
  \label{eq:flc_rate}
\end{equation}
Conversely, in the \emph{empty loss cone} limit, $\delta
J_\mathrm{orb} \ll J_\mathrm{LC}$, there is no way back from the loss
orbits and the situation can be described as a genuine diffusion in
$J$-space. At a given energy, the star density in $J$-space gradually
goes to zero as $J_\mathrm{LC}$ is approached from above. This
negative gradient controls the diffusive flux of stars to the lethal
loss orbits. Analytical treatment of this regime is far beyond the
scope of this paper so we refer the interested reader to the
above-mentioned previous studies and turn to a description of our
MC approach to the problem.

\subsection{Implementation of loss cone effects}
\label{sec:MC_LC_sim}

A reliable determination of the tidal disruption rate requires for the
numerical simulation of the relaxation process a resolution $\delta
J_\mathrm{num} < J_\mathrm{LC}$ in the empty loss-cone regime and
$\delta J_\mathrm{num} = \delta J_\mathrm{orb}$ in the full loss-cone
regime. The latter case could be treated by use of
Eq.~\ref{eq:flc_rate} as a quick shortcut but the former constraint
cannot be circumvented as easily. Unfortunately, whereas simulation of
``normal'' relaxation imposes a value of the numerical deviation angle
per step, $\delta \theta_\mathrm{step}$ sufficiently smaller than
$\pi$ ($\delta \theta_\mathrm{step} \simeq \pi/2\sqrt{f_{\delta t}}
\simeq 0.1\pi$, see Eqs.~7 and
of 10 paper~I), resolution of the (empty) loss cone region is
not attained unless $\delta \theta_\mathrm{step} < \theta_\mathrm{LC}
\ll \pi$! Furthermore, a foolproof approach, not relying on a
clear-cut {\it a priori} distinction between ``full'' and ``empty''
regimes, would necessitate to reduce $\delta \theta_\mathrm{step}$ to
the tiny ``elementary'' orbital $\delta \theta_\mathrm{orb}$ step with
a corresponding $\delta t_\mathrm{step} = P_\mathrm{orb} \simeq
\ln(\gamma N_\ast)N_\ast^{-1} T_\mathrm{rel}$, thousands of times
smaller than the desired $\delta t_\mathrm{step} \simeq f_{\delta
  t}T_\mathrm{rel}$! Although \citet{Shapiro85} was able to
attribute such tiny $\delta t$ only to those particles orbiting close
to (or inside) the LC, hence preventing too drastic a code slowing
down, such a feature doesn't fit in any straightforward way into
H\'enon's scheme. To mention but one impediment, the need of devising
time steps that depend only on the super-star's radial rank would impose
$\delta t\simeq P_\mathrm{orb}$ for a large fraction of super-stars.

The simple structure of our code -- mainly consisting in successive
2-super-star interaction steps -- having proved to be both easy to
grasp conceptually and reliable when applied to relaxational and
collisional simulations, we introduced loss cone effects in a way that 
required the least modifications. 

\begin{figure}
  \resizebox{\hsize}{!}{
    \includegraphics{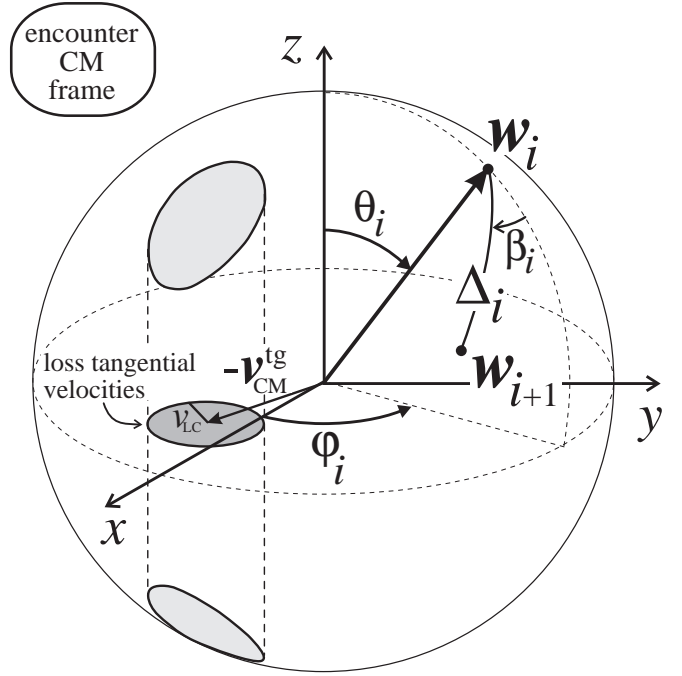}
    }
  \caption{ Geometry of one random walk step on the velocity
    sphere in the encounter reference frame. $\Delta_i$ is the
    adaptive $i^\mathrm{th}$ step, $\beta_i$ a random angle,
    $\vec{w}_i$ the particle's velocity after step $i-1$ and
    $\vec{w}_{i+1}$ its velocity after step $i$. Velocities with
    tangential component pointing in the shaded disk correspond to
    disruption orbits, i.e. with $v^\mathrm{tg} \le v_\mathrm{LC}
    \equiv J_\mathrm{LC}/R$
    in the cluster reference frame. $\vec{v}_\mathrm{CM}^\mathrm{tg}$
    is the tangential component of the pair's center of mass velocity.}
  \label{fig:lc_rw_sphere}
\end{figure}

Let's consider a single step. If the encounter was a collision, we
only need to test whether each surviving super-star entered the LC
through the interaction and to disrupt it in such a case. Indeed,
collisions are not to be refined into more elementary processes. On
the other hand, after a gravitational super-encounter has been
computed, with deflection angle $\delta \theta_\mathrm{step}$ in the
encounter reference frame, each surviving super-star is examined for
tidal disruption in turn by simulating its random walk (RW) in
$J$-space during $\delta t_\mathrm{step}$. In MC spirit, we
estimate typical ``representative'' for the diffusion angle during a
single orbit, $\delta\theta_\mathrm{orb}$ by scaling down $\delta
\theta_\mathrm{step}$ to orbital time,
\begin{equation}
  \delta\theta_\mathrm{orb} \equiv 
  n_\mathrm{orb}^{-1/2}
  \delta \theta_\mathrm{step} \mbox{\ \ with\ \ } 
  n_\mathrm{orb} = \frac{\delta t_\mathrm{step}}{P_\mathrm{orb}}.
  \label{eq:orb_defl}
\end{equation}
Let $\vec{w}$ be the super-star's velocity vector in the encounter
frame. We decompose the step $\delta t_\mathrm{step}$ into a random
walk of the tip of $\vec{w}$ on a sphere with fixed $w=\|\vec{w}\|$
radius, starting at its initial direction. A brute force
implementation would require up to $n_\mathrm{orb}$ steps of angular
size $\delta\theta_\mathrm{orb}$, each one followed by a test for
entry into the LC (Eq.~\ref{eq:J2LC}). The number of orbits per
$\delta t_\mathrm{step}$ typically ranging from $10^3$ to $10^6$, such
a procedure turns out to be extremely inefficient, requiring a huge
number of operations to detect only a few tidal disruptions, even if
super-stars with initial velocities pointing too far from the LC are
filtered out\footnote{Actually, as $\delta \theta_\mathrm{step}
  \propto \sqrt{\delta t_\mathrm{step}} \propto \sqrt{f_{\delta t}}$,
  the number of super-stars to be tested for entry into the LC per
  (mean) $\delta t_\mathrm{step}$ scales roughly as $\delta
  \theta_\mathrm{step}^2 \propto f_{\delta t}$, with $n_\mathrm{orb}
  \propto f_{\delta t}$ steps in each random walk. As the number of
  $\delta t_\mathrm{step}$ needed to simulate the cluster's evolution
  for a given physical duration is $\propto f_{\delta t}^{-1}$, the
  total number of RW steps scales as $\propto f_{\delta t}$ and the
  code gets {\bf slower} for larger time steps!}. Fortunately, the
burden can be lighten enormously through use of \emph{adaptive} RW
steps. Indeed, $n$ individual steps of length $\delta$ with random
relative orientation are statistically nearly equivalent to a single
``meta'' one of length $\Delta = \sqrt{n} \delta$,\footnote{More
  precisely, for planar RW, the length of the surrogate ``meta''-step
  should be chosen according to a Gaussian distribution with $n\delta^2$
  variance.} as long as $\Delta$ is sufficiently smaller than the
distance to the LC, to keep the risk of missing a disruption during
these $n$ RW steps at very low level. Here is the outline of the
random walk procedure:
\begin{enumerate}
\item {\bf Preparation.} The orbital period is integrated using
  Gauss-Chebychev quadrature and $\delta \theta_\mathrm{orb}$ is
  deduced from Eq.~\ref{eq:orb_defl}.
\item {\bf Initialization.} The initial angular coordinates
  $(\phi,\theta)$ of $\vec{w}= (w^x,w^y,w^z)$ are computed. We set a
  variable $L_2$ to the total quadratic deflection angle to be covered
  during $\delta t_\mathrm{step}$, $L_2 \leftarrow
  \delta\theta_\mathrm{step}^2$.
\item {\bf LC test.} If $v^\mathrm{tg} = \sqrt{
    (v_\mathrm{CM}^x+w^x)^2 + (v_\mathrm{CM}^y+w^y)^2 } \le
  v_\mathrm{LC} \equiv J_\mathrm{LC}/R$, the super-star has entered the
  loss cone and is disrupted. Otherwise, we proceed to the next step
  of the procedure. We recall that $\vec{v}_\mathrm{CM}$ is the
  velocity vector of the pair's center of mass in the cluster
  reference frame.  It is considered constant during the RW process.
  \label{enum:lc_test}
\item {\bf Completion test.} If $L_2 \le 0$, the random walk is
  over. We break from the RW loop, the super-star left unaffected. 
\item {\bf RW step.} A new (meta-)step is realized. First its
  amplitude is set according to
  \begin{equation}
    \Delta = \max \left( \delta \theta_\mathrm{orb}, \min \left(
        \Delta_\mathrm{max}, \Delta_\mathrm{safe}
        ,\sqrt{L_2} \right) \right),
  \end{equation}
  where $\Delta_\mathrm{max} \simeq 0.1\pi$ and $\Delta_\mathrm{safe}
  = c_\mathrm{safe} ( v^\mathrm{tg}-v_\mathrm{LC} )/w$ with
  $c_\mathrm{safe} \simeq 0.2$--$0.5$. This relation ensures that
  meta-steps get progressively smaller, down to the ``real''
  individual $\delta \theta_\mathrm{orb}$ when the loss cone region is
  approached during $\vec{w}$-RW. Then the (meta-)step direction on
  the sphere is set by an random angle, $\beta$, with uniform
  $[0,2\pi[$ deviate (see Fig.~\ref{fig:lc_rw_sphere}). This
  determines a new orientation $(\phi,\theta)$ for $\vec{w}$. The
  remaining quadratic path length is updated, $L_2 \leftarrow L_2 -
  \Delta^2$. The loop is closed by branching back to
  point~\ref{enum:lc_test}.
\end{enumerate} 

To conclude this section, we highlight some shortcomings in our
treatment of the LC. Our procedure amounts to examining whether tidal
disruption occurs during the fine-grained diffusion process
numerically represented by a single super-encounter. Thus, as long as
``normal'', non-LC relaxation is concerned, the super-encounter and
the explicit RW are two statistically equivalent descriptions of the
particle's evolution during $\delta t_\mathrm{step}$. But only if the
RW process leads into the LC, is the particle's $J$ modified as this
is needed to determine the outcome of the tidal interaction. Its
energy isn't modified accordingly because energy conservation would be
violated if some energy change were applied to the super-star without
being balanced by an opposite modification for the other super-star
that took part to the super-encounter\footnote{Conversely,
non-conservation of angular momentum doesn't show up explicitly for
the contribution of any super-star to the total $J$ is always zero, by
spherical symmetry! However, there is a risk that such ``hidden''
non-conservations of $J$ may reflect in the distribution of
ellipticities by introducing some nonphysical feature in it.}.  The
main risk is the introduction of some bias in the $E$-distribution of
stars that endured partial tidal disruption.  Furthermore, if the
super-star survived the RW, we give it back the post-super-encounter
orbital quantities. Hence, there a possibility that it will be left
lying in the LC with no regards to its empty/loss nature! This means
that the DF as represented by the code is probably not accurate in the
LC region.

A possible cure to these problems would be to eliminate the
super-encounter phase and to perform a symmetric RW for both super-stars
at the same time.  Unfortunately, this is not so easy for they do not
share a common $n_\mathrm{orb}$. Also, consistency would dictate to
start the random walk with the orbital properties of the super-star
(which determine $n_\mathrm{orb}$, for instance) \emph{before} the
super-encounter. However, to save computing time, the RW's initial
conditions are set to the orbital state modified by the
super-encounter, as this spares an extra computation of the peri- and
apocenter distances which are needed both to compute $P_\mathrm{orb}$
and to select a radial position on the new orbit. Quite unexpectedly,
tests have demonstrated that this trick doesn't introduce any
significant change in the cluster's evolution (most notably, the BH's
growth rate)\footnote{To be fair, the gain in speed is also quite
  modest, as most of computing time is spent in the orbital position
  selection procedure.}.

\begin{figure}
  \resizebox{\hsize}{!}{
    \includegraphics{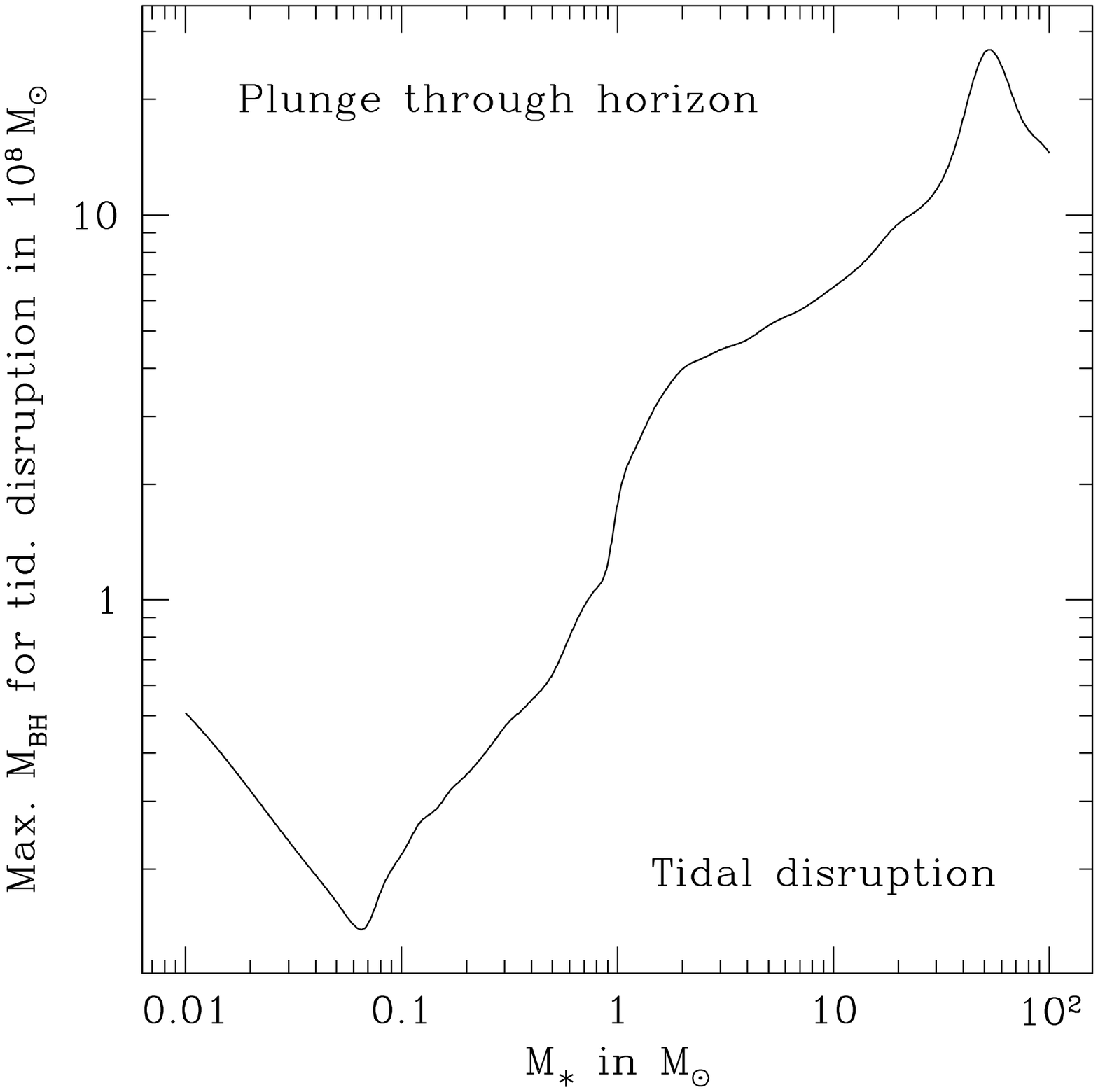}
    }

  \caption{ Maximum mass of the central BH for tidal disruption, as a
  function of the mass of of the MS star,
  $M_{\mathrm{BH}}^{(\mathrm{max})} = 1.6\times10^8\,M_\odot
  \left(\frac{M_\ast}{M_\odot}\right)^{-1/2}
  \left(\frac{R_\ast}{R_\odot}\right)^{3/2}
  \left(\frac{R_\mathrm{plunge}}{R_\mathrm{S}}\right)^{-3/2}
  \left(\frac{\beta_\mathrm{h}}{0.8}\right)^{-3/2}$. We assumed
  $R_{\mathrm{plunge}}=R_{\mathrm{S}}$ and $\beta_\mathrm{h}= 0.8$ and
  used $M_\ast$--$R_\ast$ relations from realistic models of MS stars
  \citep{SSMM92,MMSSC94,CDSBMMM99,CB00}.}

  \label{fig:Mlim_plunge}
\end{figure}

In our description, we neglected the fact that if the BH is massive
enough, its Schwarzschild radius $R_\mathrm{S} = 2GM_\mathrm{BH}/c^2$
can exceed $R_\mathrm{disr}$ for stars with a given structure so that
they will be swallowed by crossing the horizon without being
disrupted. For a star with solar mass and radius, this will happen for
$M_\mathrm{BH} > 1.6\times 10^8 M_{\sun}$ while, for giants with $M=1
M_{\sun}$ and $R=100 R_{\sun}$, only an unrealistic BH with
$M_\mathrm{BH} > 1.6\times 10^{11} M_{\sun}$ would be large enough to
prevent disruptions from happening. In Fig.~\ref{fig:Mlim_plunge}, we
plot, as a function of the mass of the MS star, the maximum BH mass
for which tidal disruption can occur.  Note, however, that assuming
$R_\mathrm{plunge} = R_\mathrm{S}$ may be an underestimate. Indeed, a
particle with negligible energy at infinity would be pulled into the
BH on a no-return in-spiral orbit by relativistic effects if its
specific angular momentum is lower than
$J_{\mathrm{min}}=4GM_{\mathrm{BH}}c^{-1}$, as the effective potential
does not have high enough a centrifugal rise.  This critical $J$ value
corresponds to a parabolic orbit with pericenter separation
$R_{\mathrm{plunge}} \stackrel{\mathrm{def}}{=} d_{\mathrm{min}}=
4R_{\mathrm{S}}$ in Newtonian mechanics
\citep[ Sec.~12.3]{ST83}. This should be used as the effective radius of the direct
plunge sphere provided tidal disruptions occurring on such relativistic
in-spiral orbits do not lead to observable accretion events, i.e. most
of the stellar gas stays on in-spiral trajectories, which seems
unlikely given the huge spread in orbital energy of the
post-disruption gas elements \citep{Rees88}. An easy modification of
the code allows to account for direct plunges but, for the sake of
comparison with results from the literature, they were not treated in
any simulations presented here.

In the present version of the code, we assume that each time a star
enter the disruption sphere, it is completely shredded to gas and that
all this gas is immediately accreted onto the central BH. Treating the
accretion process as being {\em instantaneous} is certainly a good
approximation when the mean time between successive disruptions (of
order $10^4$\,years in present-day galaxies) is much longer than the
time scale of individual accretion events (a few months to a few
years). When this is not the case, one may assume that the gas piles
up in some circum-BH reservoir, waiting to be accreted at a later
time when the disruption rate has decreased and/or the increased BH
mass allows a shorter accretion time (see models {\em \`a la}
\citealt{MCD91} in Sec.~\ref{subsec:collgalmod}).

On the other hand, assuming {\em complete accretion} probably
leads to an overestimate of the tidal feeding rate because, due to the
huge spread in the energy of debris, only 50\,\% of the stellar gas is
left bound to the BH just after a complete tidal disruption \citep[][
among others]{Rees88,Fulbright96}. Furthermore, when the leading
extremity of this bound gas stream comes back to pericenter, it
collides with slower moving material and shocks to such a high thermal
energy that of order half of the bound gas may eventually get unbound
\citep{ALP00}. Consequently, in future works, we should assume that only a
fraction $\epsilon_{\mathrm{accr}}=25-50$\,\% of the tidally produced
gas is accreted, but, to be consistent with other cluster simulations
from the literature, all results reported here were obtained with
$\epsilon_{\mathrm{accr}}=100$\,\%. 

Finally, the assumption of {\em complete disruption} is also an
over-simplification, as hinted to by, e.g., \citet{Fulbright96} who
showed that the transition regime between no damage and full
disruption spans $\beta\simeq 1 \rightarrow 3$ for $n=3$
polytropes. Real MS stars with masses $\ge1\,M_{\odot}$, not to
mention giants, are even more concentrated than $n=3$ polytropes so
that there is an important range of pericenter distances for which
envelope striping, rather than complete disruption would result. Other
non-disruptive tidal effects like spin-up \citep{AK00} are also of
observational interest for the center of our Galaxy and we plan to
extend the abilities of our code in order to be able to keep track of
such ``tidally perturbed'' stars that can amount to an appreciable
fraction of the inner stellar population \citep{AL01}.

\section{Other additions and improvements}
\label{sec:imp}

\subsection{Stellar evolution}
\label{subsec:se}

Stellar evolution (SE) is, in principle, an important ingredient to
incorporate in nuclei simulations. For a typical IMF, of order 40\,\%
of the Zero-Age MS (ZAMS) mass is lost from the stars in the first $10^{10}$~years,
so SE is potentially one of the dominating source of fuel for the
BH. Also, how stars are affected by relaxation, collisions and
tidal disruptions obviously depends on their masses and radii. For
example, compact remnants resist disruptive events and, with the
help of mass segregation, may come to dominate the central
regions. Whether or not larger and larger stars may be formed through
successive mergers also depends crucially on the relative time scales of
stellar evolution and collisions.

For the time being, our treatment of SE is simple-minded and
straightforward. We assume that a star is ``born'' on the ZAMS and
keeps the same mass and radius during its MS life which is of duration
$T_{\mathrm{MS}}$. We use the relation $T_{\mathrm{MS}}(M_{\ast})$
given by \citet{BFBC93}. When it leaves the MS, this star is
immediately turned into a compact remnant, according to the following
prescription \citep{MEG00}. All progenitors with masses lower than
$8\,M_{\sun}$ become $0.6\,M_{\sun}$ white dwarfs, those with masses
$8$--$30\,M_{\sun}$ become $1.4\,M_{\sun}$ neutron stars and those
with larger masses become $7\,M_{\sun}$ BHs. Part of the emitted gas
is accreted on the central MBH and the remaining is ejected from the
cluster.  This simplistic relation between the ZAMS mass of a star and
the final product of its evolution mainly reflects the lack of a
strong set of observational constraints or theoretical predictions in
this domain.  In any case, it is known that the
ZAMS\,$\longrightarrow$\,remnant relation strongly depends on metallicity,
if only because stellar winds do \citep{Maeder92}. All in all, it
appears to us that these aspects of SE are probably a
main source of uncertainties affecting the prediction of stellar
dynamical mechanisms in which remnants take an important part.

SE introduces a new time scale, namely
$T_{\mathrm{MS}}$ in the present implementation. To resolve it
correctly, we impose the time step $\delta t(R)$ to be smaller than a
fraction $f_{\delta t}^{\mathrm{(SE)}}$ (typically 0.05) of the
minimum of $T_{\mathrm{MS}}$ as evaluated in each cell of the same
radial mesh we use to estimate $T_{\mathrm{rel}}(R)$ and
$T_{\mathrm{coll}}(R)$. But, contrary to relaxation and collisions, in
the absence of a strong initial mass segregation, there is no reason
for this time-scale to increase with increasing $R$.  Consequently,
when SE proceeds faster than other processes, it
imposes (nearly) the same, very short $\delta t$ to all super-stars and
we loose the advantage of $R$-dependent $\delta t$. In the 
simulations we have performed so far with SE included,
we assumed a unique initial episode of star formation a $t=0$ so that,
as soon as high mass stars have been turned into remnants, the slowing
down due to stellar evolution ceases and the total CPU time is only
increased by a factor of a few. A more fearsome performance decline
will result if some form of continuous stellar formation is simulated
or if the red giant phase has to be resolved as well.

\subsection{Particle doubling}

To maintain a high resolution in the late evolutionary stages of a
highly collisional, disruptive or evaporative cluster, we resort to
particle doubling. When the number of remaining super-stars has
reached half the initial number, every super-star is split
into two copies with the same orbital and stellar properties. In the
first stage of the procedure, both copies are left at the same
position $R$ where their ``parent'' was. Then, we pick each super-star
in turn, in random order, and place it at a random position on its
orbit, in a way identical to what is done at the end of a normal
evolutionary step. In that way, we minimize the risk of maintaining
potentially harmful correlations between super-stars descending from a
common ancestor. Of course, after particle doubling, the number of
stars represented by each super-star has to be divided by 2. Some
cluster models (like the one set according to DS82 model E, see below)
go through several episodes of particle doubling. Implementing proper
book-keeping was the main difficulty with this new, otherwise
straightforward, feature.

\subsection{Miscellaneous}
\label{subsec:misc}

Various minor improvements have also been recently added to the
code. For instance, in order to ensure that the orbital parameters
($E$ and $J$) and positions of the super-stars are given time to adapt
to the (supposedly adiabatic, see Sec.~\ref{subsec:adiab})
modification of the potential, we force time steps $\delta t(R)$ to be
smaller than some fraction $f_{\mathrm{evap}}$ of the evaporation
time, $T_{\mathrm{evap}} \stackrel{\mathrm{def}}{=} M_{\mathrm{cl}}(
\mathrm{d}M_{\mathrm{cl}}/\mathrm{d}t )^{-1}$ where $M_{\mathrm{cl}}$ 
is the stellar mass of the cluster, and smaller than some fraction
$f_{\mathrm{int}}$ of the ``intern mass evolution'' time
$T_{\mathrm{int}}(R) \stackrel{\mathrm{def}}{=} M_{\mathrm{int}}(R)(
\mathrm{d}M_{\mathrm{cl}}/\mathrm{d}t )^{-1}$ where $M_{\mathrm{int}}(R)$ 
is the total mass interior of $R$. Typically, values around $0.01$ are
used for $f_{\mathrm{evap}}$ and $f_{\mathrm{int}}$.

Also, in addition to the usual test we perform each time a particle
has to be evolved, we periodically check for all the super-stars to be
bound. This is an iterative procedure because if, during the first
pass, we detect super-stars that are unbound, we remove them from the
system and this may unbound other particles.

\section{Test simulations}
\label{sec:tests}

\subsection{Adiabatic adaptation of the star cluster to the growth of a central black hole}
\label{subsec:adiab}

\begin{figure*}
  \resizebox{\hsize}{!}{ \includegraphics{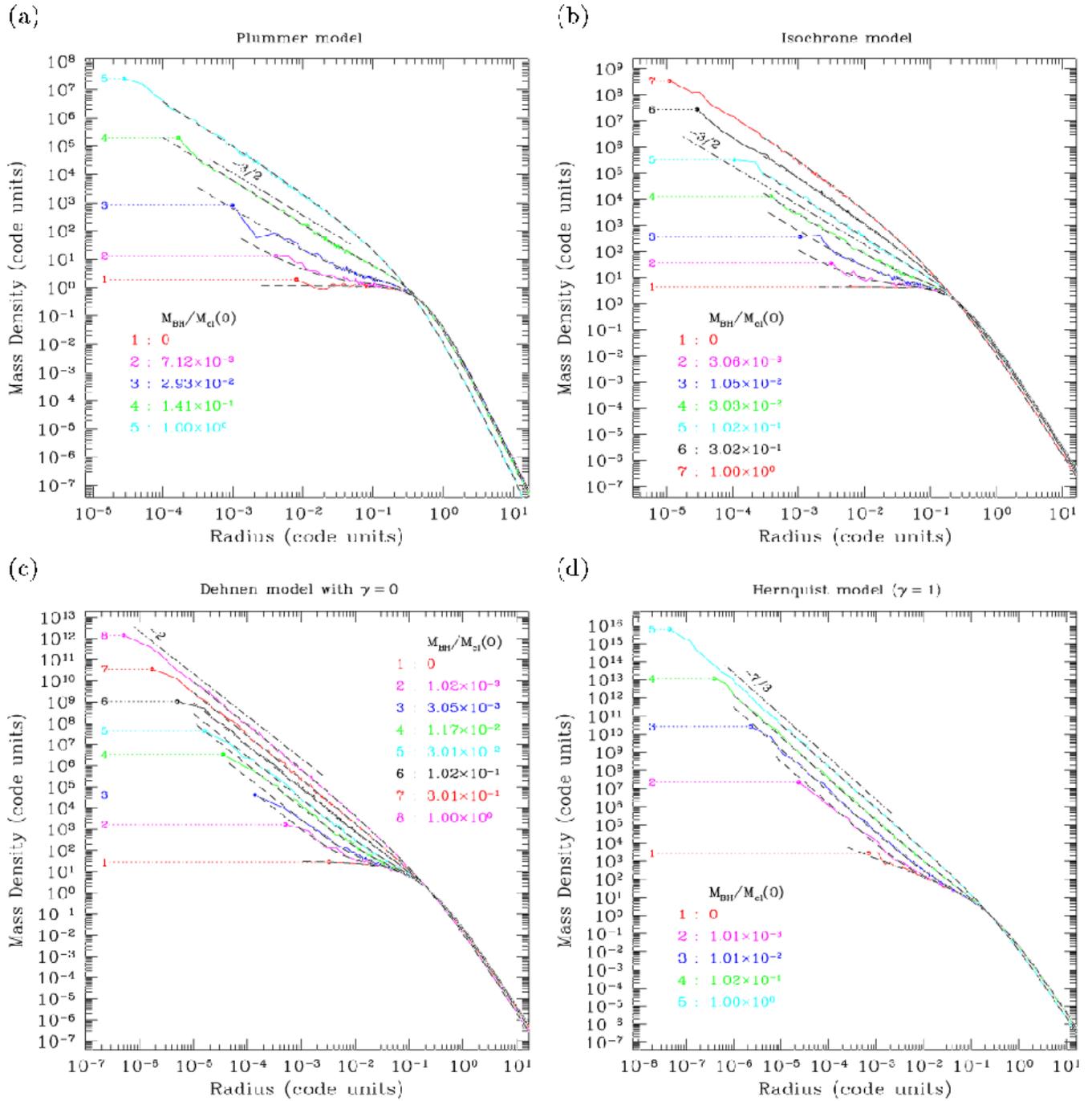} }

  \caption{ Adiabatic growth of a central BH in various cluster
  models. Evolution of the stellar density. Jagged solid lines are
  results of our MC simulations with $10^6$ super-stars. Smooth dashed
  lines are theoretical predictions based on the conservation of
  angular momentum and radial action. They have been computed with a
  code provided by G.~Quinlan \citep{QHS95}. The dot-dashed line
  segment indicates the asymptotic cusp slope from
  Eq.~\protect\ref{eq:gamma_adiab}. It applies for
  $M_{\mathrm{BH}}<M_{\mathrm{cl}}$. {\bf a)} Plummer model. {\bf b)}
  Isochrone model. {\bf c)}  $\gamma$-model with $\gamma=0$.  {\bf d)} Hernquist
    model.  The agreement between the MC results and the
  theoretical predictions is excellent.}

  \label{fig:adiab_dens}
\end{figure*}

\begin{figure*}
  \resizebox{\hsize}{!}{ \includegraphics{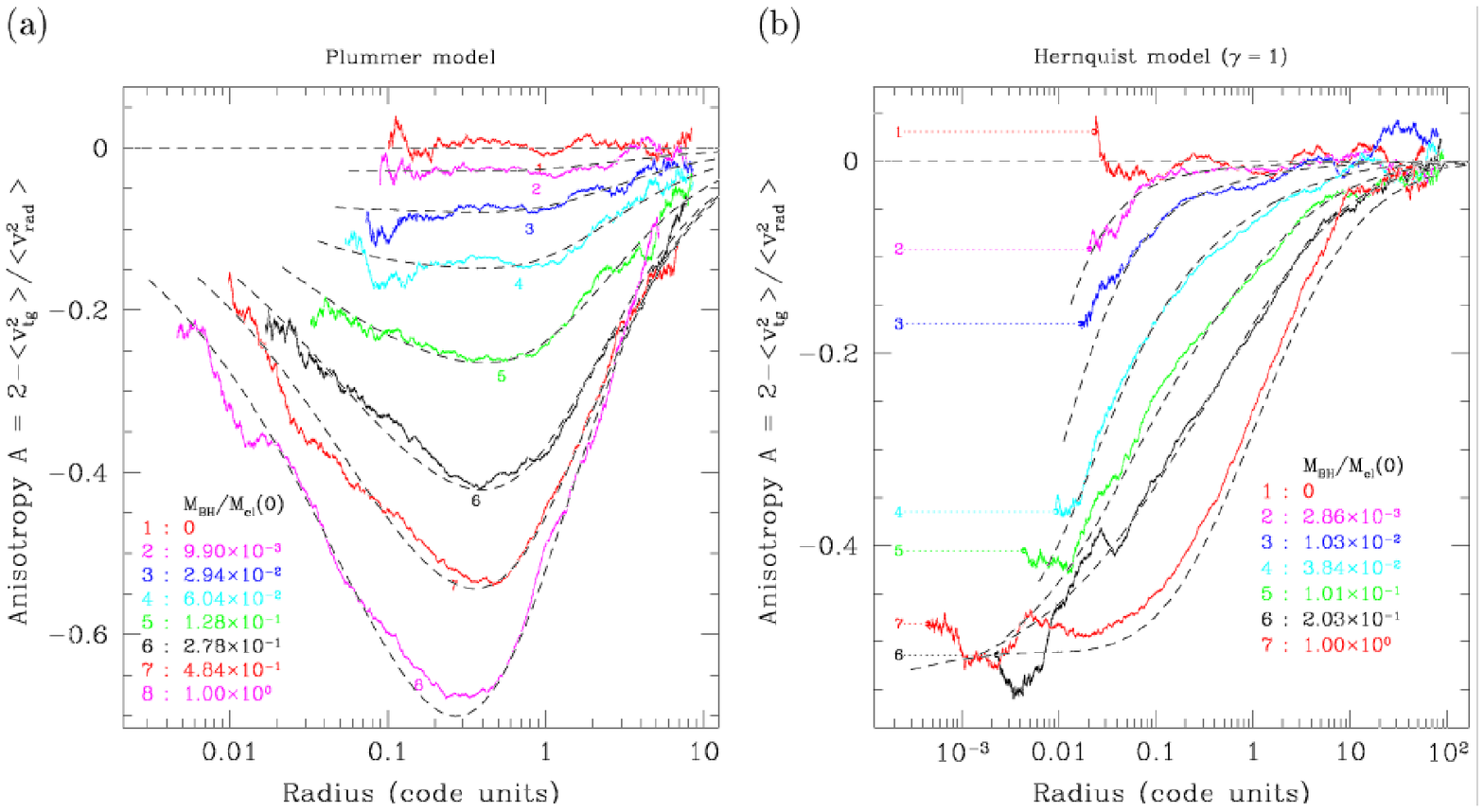} }

  \caption{ Adiabatic growth of a central BH in, {\bf a)} a Plummer
  model and, {\bf b)} a Hernquist model. Evolution of the velocity
  anisotropy. Solid lines are our results, dashed
  lines are theoretical predictions from the code of
  \citet{QHS95}. For the sake of display clarity, snapshots selected
  here are different from those in Fig.~\ref{fig:adiab_dens}. Our
  curves have been smoothed with a sliding averaging procedure. To
  cover a larger range in radius, the average is done over a smaller
  number of super-stars at small and large radii than at intermediate
  positions. Given the high level of noise in the MC data, the
  agreement with Quinlan's predictions is very satisfactory until
  $M_{\mathrm{BH}}$ grows past $0.8 M_{\mathrm{cl}}$. From this time,
  the tangential anisotropy in the outer parts of our models fails to
  increase with larger BH masses (see text). }

  \label{fig:adiab_anis}
\end{figure*}
   
In some instances the central BH can grow significantly on a time
scale $T_{\mathrm{BH}}$ much longer than the cluster's dynamical time
but still much shorter than relaxation time. Such a hierarchy
naturally occurs if a substantial amount of gas is flowing into the BH
from outside the nucleus. Quick BH growth can also happen if mass
lost my stars, either due to normal stellar evolution (in a young
cluster), or to disruptive collisions (in a very dense cluster), is
efficiently accreted on the BH.

As a consequence of the slow modification of the potential, the shape
of stellar orbits evolve while conserving adiabatic invariants, i.e.
the angular momentum $J$ and the radial action $I_R$
\citep{Young80,BT87}. Correspondingly, the density profile of stars
around the BH and their velocity distribution are modified.
Characteristics of the resulting stellar profiles have been worked out
for various initial clusters, either semi-analytically, using the
conservation of the DF when expressed as a function of adiabatic
invariants
\citep{Young80,LG89,CB94,Cipollina95,QHS95} or by means of
$N$-body simulations \citep{SHQ95,LA00}. 

These studies show that a power-law cusp develops inside the influence
sphere of the BH, of radius $R_{\bullet}$, in which
$GM_{\mathrm{BH}}/R$ exceeds the original velocity dispersion of the
stars. According to \citep{QHS95}, if the initial stellar cluster is
isotropic, presents a density cusp $\rho \propto
R^{-\gamma_\mathrm{i}}$ with $\gamma_\mathrm{i}\ge 0$ and a DF
diverging near $E=\phi(0)$ like $f(E)\propto(E-\phi(0))^{-n}$, then
the final density cusp has an exponent
\begin{equation}
\label{eq:gamma_adiab}
  \gamma_\mathrm{f} = \frac{3}{2} + n \left( 
    \frac{2-\gamma_\mathrm{i}}{4-\gamma_\mathrm{i}} \right).
\end{equation} 
This result only applies very close to the BH if its mass is larger
than the mass of the initial stellar core; the cusp may be steeper at
intermediate distances, $R_{\mathrm{trans}} <R< R_{\bullet}$ with
$R_{\mathrm{trans}}=R_\mathrm{c}^2/R_{\bullet}$ (\citealt{LG89,CB94},
see also \citealt{LA00}). Another key feature is the development of
noticeable \emph{tangential} anisotropy in the central regions. In
models with analytic cores (i.e. with $(\rho(0)-\rho(R)) \propto R^2$
near the center), this anisotropy, although it is caused by the
central BH, does not actually appear in the center itself where
isotropy is conserved \citep{GB84,QHS95}.

We have performed simulations of the adiabatic growth of a central BH
in a variety of cases. In addition to the traditional Plummer model,
we adopted the same set of models as \citet{QHS95}. These are the
isochrone cluster \citep{Henon59,Henon60,BT87}, which has an analytic
core, and three '$\gamma$-models'
\citep{Dehnen93,Tremaine94} whose density profile is
\begin{equation}
  \rho_{\gamma}(R) = \frac{3-\gamma}{4\pi} 
  \frac{M_{\mathrm{cl}}R_{\mathrm{b}}}{R^{\gamma}(R+R_{\mathrm{b}})^{4-\gamma}}
\end{equation}
where $R_{\mathrm{b}}$ is the break radius.
The used $\gamma$ values are 0, 1 \citep{Hernquist90} and 2
\citep{Jaffe83}. None of these models has an analytic core.
 Eq.~\ref{eq:gamma_adiab} predicts $\gamma_{\mathrm{f}}=3/2$, $3/2$,
 $2$, $7/3$ and $5/2$ for Plummer, isochrone, and $\gamma=0$, $1$, $2$
 models, respectively
\citep{QHS95}.

To simulate the process of adiabatic BH growth, we switched off
relaxation and all the other physical processes in the MC code. The
algorithm reduces then to moving super-stars on their orbits again and
again (see Sec.~5.2 of paper~I) while $M_{\mathrm{BH}}$ is slowly
increased. The time step condition is $f_{\mathrm{int}}=0.002$ (see
Sec.~\ref{subsec:misc}).  This relatively small value is required to
get a correct evolution of the anisotropy in the outer parts of the
cluster. With larger time steps, the particles at large radii react
too impulsively to the BH's growth and their orbits tend not to
develop enough tangential anisotropy or even to become radially
dominated. Note, however, that this problem only occurs when the BH's
mass is larger than half the mass of the stellar cluster and that the
density profile appears to be unaffected by this even for
$f_\mathrm{int}=0.01$.

In Fig.~\ref{fig:adiab_dens}, we compare our results with the output
of the code written by 
\citet{QHS95} and kindly provided by van der Marel. This code makes
explicit use of the conservation of adiabatic invariants to determine
the structure of the BH-embedding cluster and we can regard its
results as secure predictions. As can be seen on these diagrams, the
MC code behaves very nicely in this regime. Given the numerical noise
to be expected from such a method, the density profiles are deemed to
be in perfect agreement for all models.  In Fig.~\ref{fig:adiab_anis},
the evolution of the anisotropy profile for the Plummer and the
Hernquist model is plotted. This quantity, when determined from MC
results, suffers from a much higher statistical noise, so that a
stronger smoothing must be applied to get useful curves. Despite this
noise, it is quite clear that our results match the predictions very
well, except for the outer parts that lack some tangential anisotropy
for large $M_{\mathrm{BH}}$, as already discussed.

\subsection{Cluster models with tidal disruptions}
\label{subsec:modtiddisr}

\begin{figure}
  \resizebox{\hsize}{!}{ \includegraphics{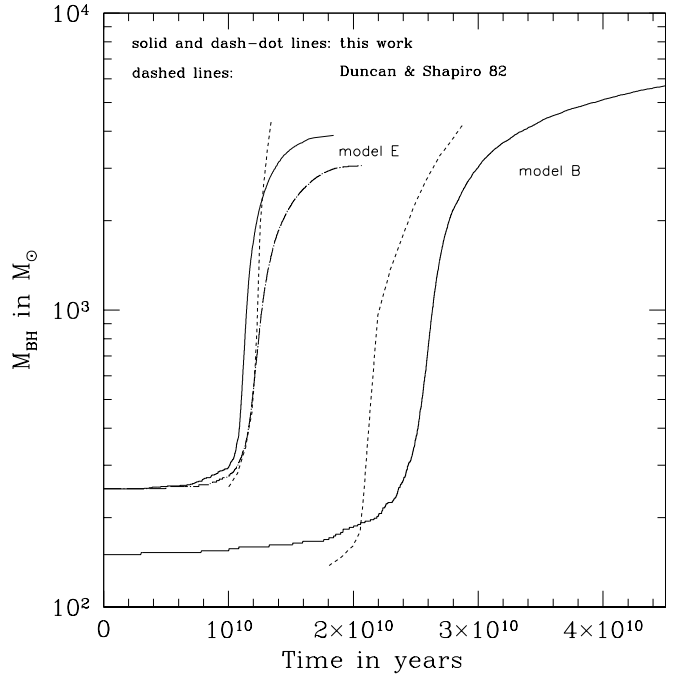} }

  \caption{ Growth of the central BH for models with initial
  conditions similar to models B and E of \citet{DS82}. Our results,
  obtained with 256k super-stars, (solid and dash-dotted lines) are
  compared with those of these authors (dashed lines). We made two
  simulations of model E. Both have been stopped when the stellar
  cluster was reduced to $500\,M_\odot$. In the first one (solid
  line), we start abruptly with a tidal radius smaller than the
  cluster which rapidly adapt to this truncation.  In the second run
  (dash-dotted line), we let the cluster adapt gently to the tidal
  truncation before we actually start the simulation by switching on
  relaxation (see text).}

  \label{fig:MBH_DS82}
\end{figure}

Another idealized regime to which many theoretical and numerical
studies have been devoted is the case of a the relaxed single-mass
spherical stellar cluster with a central BH. Collisions are neglected
but stars entering the tidal disruption region are destroyed and their
mass is added to the BH. \citet{BW76} demonstrated that the
quasi-steady state solution of the Fokker-Planck equation for this
situation corresponds to a central density cusp with $\rho\propto
R^{-7/4}$. Although these authors used an one-dimensional approach
with the energy $E$ as the only variable, more accurate numerical
integrations of the stationary FP equation in $(E,J)$ space, with a
proper account of loss cone effects, have confirmed this result
\citep{LS77,CK78}, as did evolutionary models 
\citep[][ hereafter DS83, for instance]{DS83}. 
As testified by Fig.~\ref{fig:dens_DS83I}, we reproduce this
result. This plot shows the evolution of a model with same physics and
initial parameters as model~I of DS83 and is described in more details
in the next sub-section.

A few evolutionary models have been published that are based on these
simple physical assumptions. Most were meant to explore the
possibility of forming a MBH during the core collapse of a globular
cluster\footnote{However, neglecting the role of a mass spectrum and
binary stars, they fall short of physical realism. Unless the cluster
is born with a very high velocity dispersion, $\sigma_v^2\propto M/R
\gg V_\mathrm{orb}^2$, where $M$ is the total mass, $R$ a measure of
the size of the cluster and $V_\mathrm{orb}$ a typical value for the
(internal) orbital velocity of binaries (an unrealistic assumption for
globular cluster but which may apply to models of proto-nuclei of
galaxy like those of \citet{QS90}), the binaries will delay collapse
and probably trigger core rebound before the central density is high
enough for efficient ``tidal feeding'' of a seed BH \citep[see][ for
simulations of globular clusters with primordial
binaries]{GGCM91,GS00,Giersz01,RFJ01}.}. They usually start with a
seed black hole which grows by consuming stars. To check the tidal
disruption rate given by our code, we compare the growth of the
central BH in such models with results from the literature.

Fig.~\ref{fig:MBH_DS82} shows such a comparison for models B and E of
\citet[][ hereafter DS82]{DS82} to whom we refer for the specification
of initial and boundary conditions. We have used the same setting as
these authors except that, in our computations, there is no initial
stellar cusp around the BH and that, for model~B, the BH is present
from the beginning of the simulation and not added at a later
time as done in DS82. We don't think these minor changes have any
significant effect because the initial BH amounts to only a tiny
fraction of the cluster's mass ($M_{\mathrm{BH}}=150$,
$250\,M_{\odot}$, respectively, with
$M_{\mathrm{cl}}=3\times10^5\,M_{\odot}$). The match between our
results and those of DS82 is not very good. In particular, for
model~B, the BH's growth starts at a significantly later time but
produce an object of comparable mass. However, DS82's simulations were
stopped shortly after core rebound, which does not allow a comparison
at late times. Note that the growth starts when core collapse is
sufficiently deep to bring many stars close enough to the BH to be
disrupted and that it is stopped by the fact that the disruption of
these stars, most of which have large negative energies, amounts to
heating the stellar cluster. Consequently, the temporal shift between
DS82's growth curve and ours mostly reflects that our code predicts a
longer core-collapse time, $T_{\mathrm{cc}}$. We refer to paper~I for
a discussion of this point and the large spread found in the
literature for the value of $T_{\mathrm{cc}}$.

Concerning model E, on the one hand, our value for the time of
strongest growth, again a quantity nearly coincident with
$T_{\mathrm{cc}}$, nicely agrees with DS82. Note that this cluster,
being a Plummer with a strong tidal truncation, evolves quicker and
differently than an isolated cluster, which gives more weight to this
agreement. At the end of our simulations, around 20\,Gyrs, the cluster
has nearly completely evaporated. On the other hand, the BH's growth
is steeper and stronger in DS82's simulation. There is no doubt that
it would have produced a significantly larger final BH than in our
case, had their simulation been carried on up to cluster
dissolution. The reason for this disagreement is not known to us. We
suspected that it may be linked to the fact that, in our simulation,
the remaining cluster mass is lower at all times than in DS82, which
may, in turn, be due to the way our and DS82's code cope with the
strongly out-of-equilibrium initial conditions. Indeed $\sim 10$\,\%
of all stars are initially beyond tidal radius.  In our model, the
cluster loses 17\,\% of its super-stars very quickly to adjust to the
tidal truncation. To have a better handle on this problem, we re-made
the simulation with a cluster model which was first allowed to settle
to equilibrium with its tidal truncation. To do this, we ``evolved''
it with no relaxation or any other physical process but still moving
super-stars on their orbits in the usual way. If a selected super-star
was found with apocenter beyond tidal radius, it had only a small
probability (around 0.01) to be removed at this step and was otherwise
kept (at the same position). We think that this method produces a
better initial structure in which each super-star has been given time
to react ``adiabatically'' to the enforcement of the tidal
truncation. 15\,\% of the cluster mass is lost in this procedure and
the resulting cluster also shows less evaporation during its further,
relaxation-driven, evolution. However, this does only increase the
discrepancy with DS82 concerning the final mass of the BH, see the
dash-dotted curve on Fig.~\ref{fig:MBH_DS82}. Our higher evaporation
rate is probably due to our simpler prescription for escape. We
immediately remove any super-star which gets on an orbit with
apocenter distance beyond tidal radius, regardless of its actual
position on this orbit. More realistically, DS82 allowed stars on
escape orbits to be kicked back to bound orbits. Recent works
\citep{FH00,TPZ00,Baumgardt01} made it clear that evaporation from a
cluster with a relatively low number of stars can not be regarded as
instantaneous: it takes of order one orbital time for a star to
actually leave the cluster and the probability for it to be
back-scattered onto a bound orbit is non vanishing. Whether or not
some improvement in the line of this in our evaporation prescription
would lead to a better agreement with DS82 concerning
$M_{\mathrm{BH}}(t)$ is not obvious as these two aspects may well be
uncoupled. Note that a similar mismatch in the BH's growth curve
appears in comparisons with preliminary simulations realized by
\citet{ASS01b} with a gas code (see below), but doesn't show up in
comparisons with other results obtained by
\citet{DS83} with their MC code and by
\citet{MCD91} with a direct Fokker-Planck scheme (see next subsection).

\begin{figure}
  \resizebox{\hsize}{!}{ \includegraphics{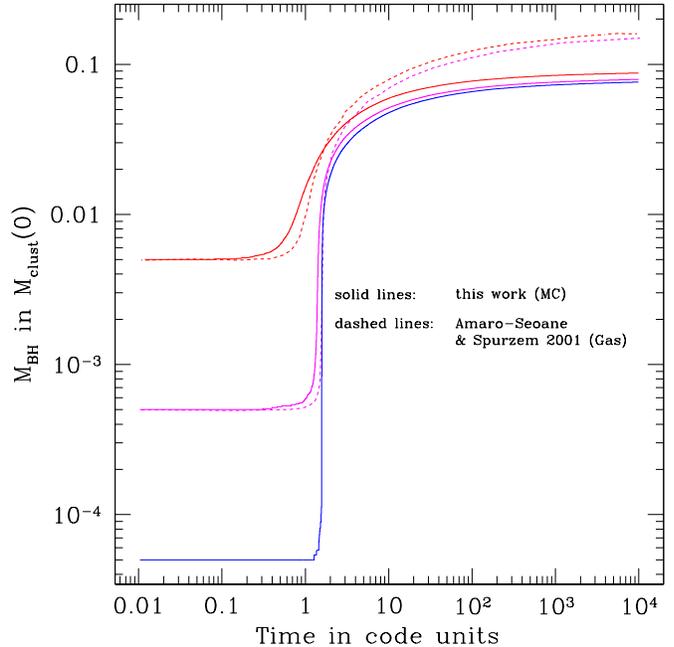} }

  \caption{ Growth of the central BH for models with initial
  conditions identical to those of \citet{ASS01b}. Our results,
  obtained with 256k super-stars, (solid lines) are compared with those
  of these authors (dashed lines).}

  \label{fig:MBH_ASS01}
\end{figure}

In Fig.~\ref{fig:MBH_ASS01}, we display the growth of the central BH
for clusters corresponding to the models used by \citet[][ hereafter
AS01]{ASS01b}. These consist of $10^5$ $1\,M_{\odot}$ stars
distributed according to a Plummer density law with a core radius of
0.707\,pc. The cluster is seeded
by a fixed central BH with an initial mass of 5, 50 or
$500\,M_{\odot}$. Only the last 2 values have been used by AS01. It is
clear that for masses as low as 5 or even 50\,$M_{\odot}$, neglecting
the motion of the BH is quite an unphysical assumption which is
required by the present limitations of numerical codes. Even if a
close agreement is not reached, our results are very similar to the
curves from AS01. In particular, we get the same phenomenon of
convergence at late times toward an unique value of
$M_{\mathrm{BH}}$. This value is however smaller by a factor of
$\sim2$ than that of AS01.

\subsection{Galactic nucleus models including collisions}
\label{subsec:collgalmod}
 
\begin{figure} 
  \resizebox{\hsize}{!}{ \includegraphics{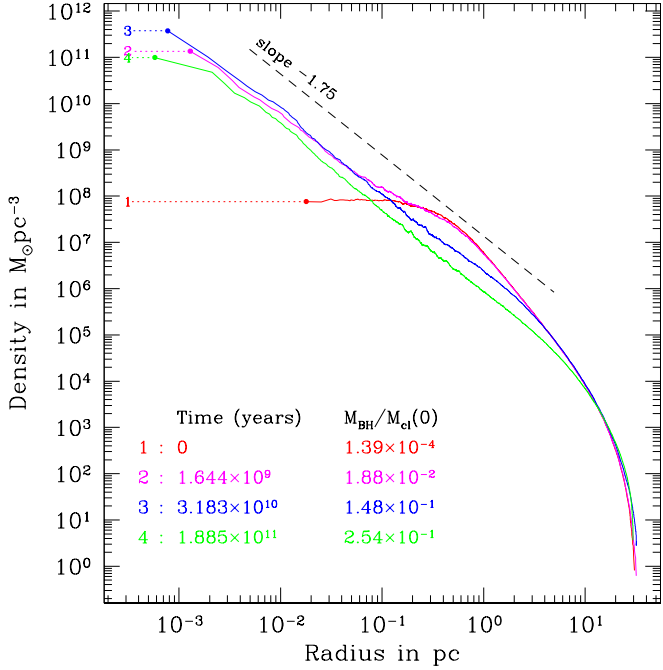} }
  \caption{ Evolution of the density profile for a cluster with
  initial conditions identical to models I/II of DS83. The initial
  number of super-stars is $10^6$. As in DS83's model I, collisions
  are not simulated. One notes the rapid development of a $\rho\propto
  R^{-1.75}$ cusp. } 
  \label{fig:dens_DS83I}
\end{figure}

\begin{figure} 
  \resizebox{\hsize}{!}{
    \includegraphics{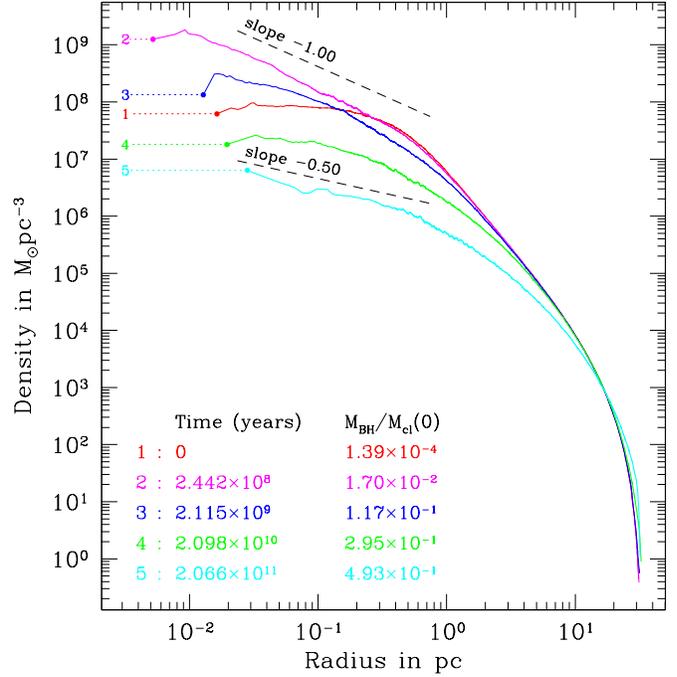}
    }
  \caption{
    Evolution of the density profile for a cluster with initial
    conditions identical to models I/II of DS83. The initial number of
    super-stars is $2\times10^6$. As in DS83's model II, collisions
    are simulated. They are assumed to be completely disruptive.
    Instead of a steep $\rho\propto R^{-1.75}$ power law, the cusp in
    the center gets milder and milder. }
  \label{fig:dens_DS83II}
\end{figure}

\begin{figure}
  \resizebox{\hsize}{!}{
    \includegraphics{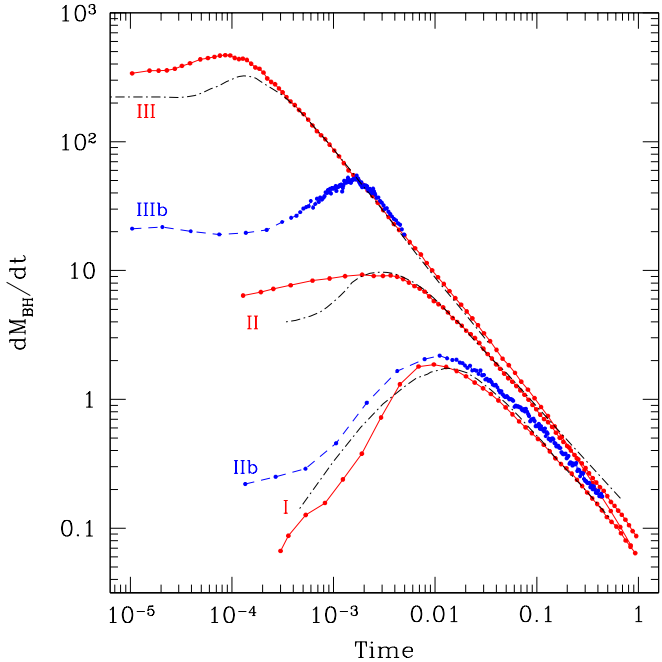}
    }
  \caption{
    Evolution of the growth rate of the central BH in clusters
    with initial conditions identical to models I, II and III of DS83.
    Dot-dashed lines are from DS83. Model I does not include stellar
    collisions. Models II and III treat them as causing complete
    disruption of stars. Solid lines with dots are our results for
    these systems. Dashed lines with dots (labeled IIb and IIIb) show
    the effects of a realistic, SPH-generated, prescription for the
    outcome of collisions which allows partial disruptions and mergers
    (see text). We used 512\,000 to $2\times10^6$ super-stars in our
    simulations. ``$N$-body'' units are used. For models I and II, the
    time unit is $1.37\times 10^{11}$\,yrs and the unit for
    $\mathrm{d}M/\mathrm{d}t$ is $2.6\times 10^{-3}\,
    M_{\sun}\mathrm{yr}^{-1}$. In model III, these units are
    $9.81\times 10^{11}$\,yrs and $5.8\times 10^{-3}\,
    M_{\sun}\mathrm{yr}^{-1}$.}
  \label{fig:DS83_BH_GrRate}
\end{figure} 

\begin{figure}
  \resizebox{\hsize}{!}{
    \includegraphics{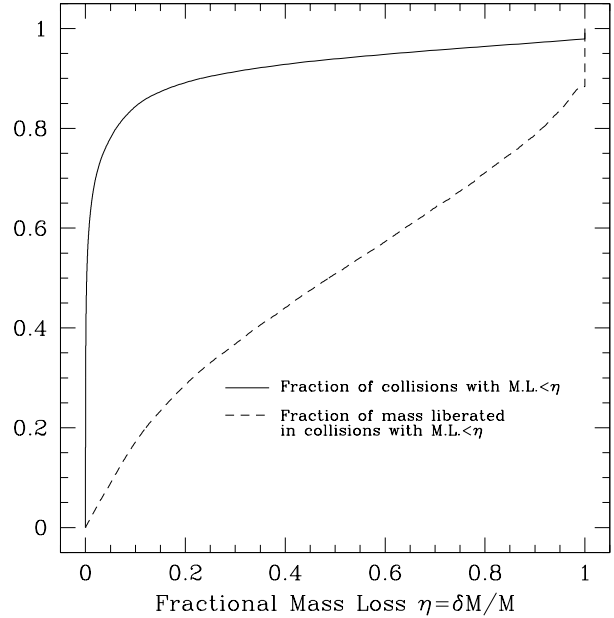}
    }
  \caption{ Cumulative distribution of the fractional mass losses in
    collisions for a simulation of model II with improved treatment of
    collisions (see text). All collisions occurring before time $T=0.1\,
    \tilde{\mathcal{U}}_\mathrm{t} = 1.37\times 10^{10}
    \,\mathrm{yrs}$ are included in this count. The solid line shows the
    number fraction of all collisions which resulted in a fractional
    mass loss lower than a given amount $\eta$. The dashed line
    indicates what mass fraction of collisionally released gas came
    from collisions with fractional mass loss lower than $\eta$.}
  \label{fig:dist_cum_dM_coll}
\end{figure}

After having checked individual aspects of the MC code in simplified
models (pure relaxation in paper~I, collisions rates in
sections~\ref{sec:coll_rate} and \ref{sec:rel_coll_rate_m1m2},
adiabatic BH growth in Sec.~\ref{subsec:adiab}\ldots), we turned to
the few published works addressing the long term evolution of dense
galactic nuclei in order to check our code's global behavior in
physical regimes more relevant to our astrophysical field of
interest. 

We first wanted to avoid the extra complication of stellar evolution
and discarded those papers which take it into account. Furthermore, by
their nature, Fokker-Planck methods can only include collisional
effects in an approximate way so that they don't allow a clear check
of this aspect of the code. Finally, $N$-body simulations
\citep{Arab97,Rauch99}, although much more realistic\footnote{In the
case of \citet{Arab97}, it is not clear, however, how reliably
relaxation processes can be simulated with a TREE algorithm.}, were
deemed too noisy to provide reliable data to compare with.

So we chose the venerable models by \citet[][ hereafter DS83]{DS83} to
conduct tests that include relaxation, tidal disruptions and stellar
collisions. DS83 studied three different models. The initial structure
is a King cluster with $W_0=8$ made of identical stars with
$M_\ast=1\,M_{\sun}$. Models I and II share the same
initial conditions: $3.6\times 10^{8}$ stars and a core radius $R_\mathrm{c} =
0.50\,\mathrm{pc}$ (the total radius is $34.7\,
\mathrm{pc}$). A seed black hole is present at the center with an
initial mass $M_\mathrm{BH}(0)=5\times 10^4\,M_{\sun}$. Model III was
devised to reach quasar-like accretion rates. It initially contains
$57\times 10^{8}$ stars, it has $R_\mathrm{c} = 0.82\,\mathrm{pc}$ and
$M_\mathrm{BH}(0)=2\times 10^6\,M_{\sun}$. Models II and III include
stellar collisions. They are assumed to be completely disruptive and
the gas they release is instantaneously and completely accreted on the
BH. We used the same initial conditions and physics but, to assess the
influence of the assumption of complete collisional destruction, we
carried out two extra simulations using our realistic, SPH-generated,
prescriptions (models IIb and IIIb).

In figures \ref{fig:dens_DS83I} and \ref{fig:dens_DS83II}, we present
the evolution of the density profile for models I and II,
respectively. The most conspicuous feature of the first figure is a
spreading central cusp with $\rho\propto R^{-7/4}$. Such a power-law
profile is reproduced here for the first time by a H{\'e}non-like
Monte Carlo method.  Fig.~\ref{fig:dens_DS83II} shows that when
disruptive collisions are introduced in our calculations, as in DS83's
model II, a much milder cusp first appears (with exponent $\sim -1$)
and progressively flattens (with exponent $\geq -0.5$). It has been
repeatedly reported that collisions strongly decrease the steepness of
the inner density profile \citep{DS83,MCD91,DDC87a,DDC87b,Rauch99}. A
slope of $\sim -0.5$ is often obtained. However, the simulations by
\citet{Rauch99} point to the establishment of a flat, cusp-less
central region, not unlike our own results. \citet{MCD91} get a strong
depletion of stars in the innermost part of the cluster, a result
which is apparently reproduced in some of Rauch's models. For lack of
resolution, there is no similar effect to be seen in our
simulations. The practical relevance of this discrepancy is probably
low, however, because the size of this rarefied zone is so small that
it would contain only a few $M_{\odot}$ in most cases even without
depletion. So the validity of a statistical treatment of such a tiny
region is highly questionable anyway. The evolution of the density
profile for model III is qualitatively similar. Interestingly, model
IIb, which incorporate realistic, partially disruptive collisions also
forms a $R^{-7/4}$ cusp, but in the much denser model IIIb, collisions
are efficient enough to reduce the exponent to a value between -1
and -0.5.

The growth rate of the BH is depicted in Fig.~\ref{fig:DS83_BH_GrRate}. The
qualitative agreement with DS83 is satisfying even though the rate we
obtain is higher by a factor of $\sim 2$ in initial phases of
collisional models. The reason of this difference is unknown to us.
The most important effect of a realistic treatment of collisional
outcome is a strongly reduced accretion rate. This is mainly due to
the fact that most collisions are grazing and consequently produce low
mass losses even for high relative velocities. Indeed, neglecting
gravitational focusing, we get
\[
  \frac{\mathrm{d}N_\mathrm{coll}}{\mathrm{d}d_\mathrm{min}} \propto
  \frac{d_\mathrm{min}}{R_1+R_2} \mbox{\ \ for\ } d_\mathrm{min} <
  R_1+R_2
\]
where $d_\mathrm{min}$ is the closest encounter distance for the
equivalent 2 point-mass problem. The cumulative distribution of the
fractional mass loss for model II is depicted in
Fig.~\ref{fig:dist_cum_dM_coll}. Actually, the average mass loss per
collision is as low as $0.08\,M_{\sun}$ despite an average relative
velocity for collisions of $v_\mathrm{rel}=8.8\,v_{\ast}$ (see
Eq.~\ref{eq:Vesc_stell}). These examples clearly demonstrate that any
incorporation of collisions in galactic nuclei dynamics must account
for partially disruptive events.

To conclude this series of tests, we turn to one of the most complete
and widely used set of simulations of the long-term evolution of dense
galactic nuclei published to date, namely the ``direct'' Fokker-Planck
integrations by
\citet[][ hereafter MCD91]{MCD91}. These authors included the following 
physics in their computations:

\begin{itemize}

\item {\bf 2-body relaxation}. 
It is treated in the standard Fokker-Planck way (for a description of
the multi-mass FP scheme see, e.g. \citealt{MC88} and references
therein). Note that, in the FP scheme, the cluster is represented as a
set of DFs, each of which represent a discretized mass class, i.e.,
stars that have all the same stellar mass.

\item {\bf Stellar collisions}. 
To get the mass loss for individual collisions, MCD91 use a
semi-analytical method derived from the procedure invented by
\citet{SS66}. It works by decomposing the stars into thin 
columns of gas parallel to the relative velocity and imposing
conservation of momentum for each, completely inelastic, collision
between a column from one star and the corresponding column of the
other star. No lateral mass, energy or momentum transport is
considered.  The MS stars are assumed to be $n=3$ polytropes with
$M_{\ast} \propto R_{\ast}$. These mass-loss rates are then averaged
over impact parameter and relative velocities to get rates that depend
only on velocity dispersion and mass ratio which allows the authors to
compute the instantaneous mass-loss rate for any mass class, due to
collisions with stars from any other (or same) mass class. The total
mass loss for a given time step and mass class is then converted into
a number of stars to be removed from the class. This is obviously
quite an inaccurate representation of the real way collisions change
the masses of individual stars. Mergers are not included in
this formalism.

\item {\bf Tidal disruptions}. 
Stars that get closer to the BH than the tidal disruption radius
are assumed to be completely disrupted and their mass is
instantaneously and fully accreted by the BH. Although our numerical
scheme is widely different, we use basically the same
assumptions, here. Hence, we refer to MCD91 and
\citet{CK78} for a description of how this is implemented in FP codes. 

\item {\bf Stellar evolution}. 
A simple prescription is used in which stars stay on the MS for
$T_{\mathrm{MS}}(M_{\ast})$ and then turn abruptly into compact
remnants (CR). No giant phase is simulated and all mass loss occurs at
the end of the MS. See MCD91 for the specification of
$T_{\mathrm{MS}}(M_{\ast})$ and the MS~$\rightarrow$~CR relation.

\end{itemize}

The initial stellar clusters are Plummer models with a core radius of
1\,pc. The total stellar mass is initially $8.291\times(10^9, 10^8,
10^7, 10^6)\,M_{\odot}$ for models of classes ``1'', ``2'', ``3'' and
``4'', respectively. The stars are initially on the MS and obey a
power-law mass spectrum, ${\mathrm{d}N_{\ast}}/{\mathrm{d}M_{\ast}}
\propto M_{\ast}^{-\alpha}$ between $0.3$ and $30\,M_{\odot}$, with
$\alpha=1.5$, 2.5 and 3.5 for cases ``A'', ``B'', ``C''. 

The cluster is seeded with a BH of mass
$M_\mathrm{BH}=10^4\,M_{\odot}$ at its center. The BH eventually
swallows all the gas lost by stars, through normal evolution,
collisions or tidal disruptions, but its growth rate is limited by the
Eddington rate $\dot{M}_{\mathrm{E}} = L_{\mathrm{E}}/(\eta c^2) =
4\pi G
\mu_{e} M_{\mathrm{BH}} m_{\mathrm{p}} / (\eta c \sigma_\mathrm{T})
\simeq
2.5\times10^{-2}\,M_{\odot}\mathrm{yr}^{-1}\,(\eta/0.1)^{-1}(M_{\mathrm{BH}}/10^6\,M_{\odot})$
where $\eta$ is the efficiency factor for conversion of mass into
radiation during the accretion process, $\mu_{e}$ is the molecular
weight per free electron of the accreted gas ($\simeq 1.13$ for solar
composition), $m_{\mathrm{p}}$ the mass of the proton and
$\sigma_\mathrm{T}$ Thomson's cross-section. A ``standard'' value of
$\eta=0.1$ is used. If the instantaneous rate of gas production from
the stars, $\dot{M}_{\mathrm{prod}}$, exceeds $\dot{M}_{\mathrm{E}}$,
only an amount $\dot{M}_{\mathrm{E}}$ accretes on the BH while the
remaining accumulates into a central ``reservoir'' --presumably an
accretion disk-- to be accreted later when $\dot{M}_{\mathrm{prod}}$
has declined below $\dot{M}_{\mathrm{E}}$. The gas is assumed to be
funneled completely and instantaneously to the center, i.e. no gas
remains in the stellar cluster or is expelled from the nucleus. The
structure of this reservoir is not resolved in the
simulations. Instead, it is assumed to be small enough to contribute
to the potential as a central point mass, exactly as the BH. However,
distributing the central mass in two components, the BH and this
reservoir, can still influence the dynamics slightly through the fact
that only the mass of the BH is used to compute the tidal disruption
radius. On the other hand, interactions between the gas reservoir and
stars are neglected (see Sec.~\ref{subsec:futurework}).

\begin{figure}
  \resizebox{\hsize}{!}{ \includegraphics{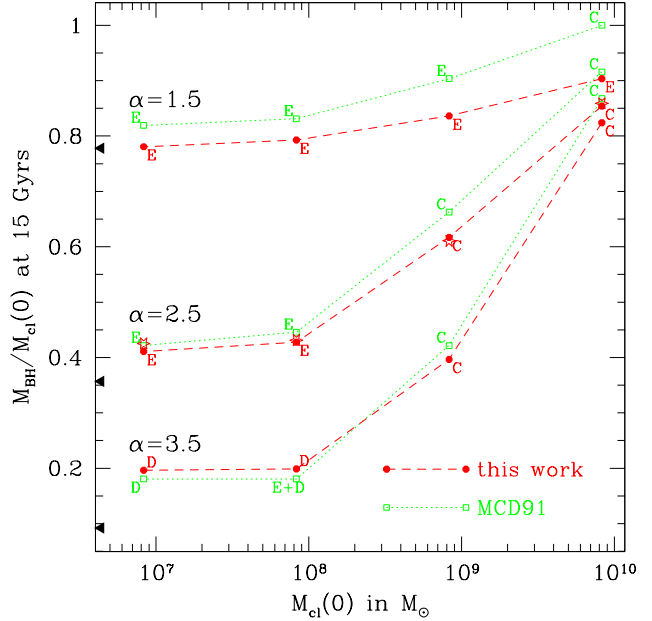} }

  \caption{ Final BH mass for all the MCD91-like models. The lines
  connect models with the same IMF slope. We compare our results
  (dashed lines) to those from MCD91 (dotted lines). Solid dots are
  for simulations with 256\,000 super-stars; the open star symbols are
  for B models with $10^6$ super-stars. The triangles on the left axis
  indicate the total fractional mass loss due to stellar evolution
  for IMF with $\alpha=1.5, 2.5, 3.5$, at an age of 15\,Gyrs. This
  corresponds to the final BH's mass expected if stellar evolution was
  the only feeding process and no star could escape the
  nucleus. Letters indicate the process whose contribution to the
  final BH's mass dominates: ``E'' stands for stellar evolution, ``C''
  for collisions and ``D'' for tidal disruptions. Most of the
  discrepancies between our results and those of MCD91 is due to the
  lower contribution of collisions (see text).}

  \label{fig:end_MBH_MCD91}
\end{figure}

\begin{figure*}
  \resizebox{\hsize}{!}{%
    \includegraphics{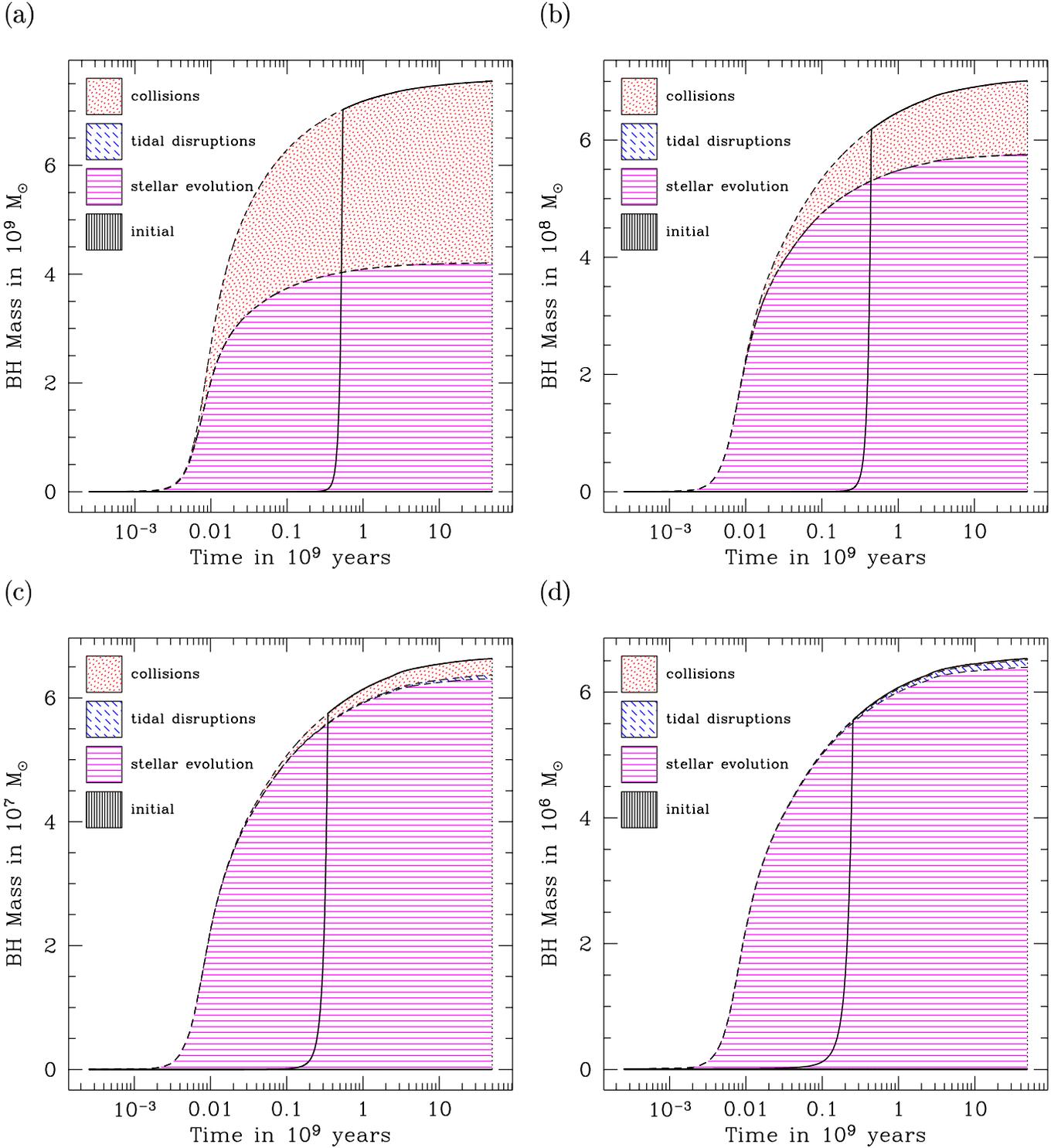}}

  \caption{ Evolution of the central mass (BH\,$+$\,gas reservoir) for
  MCD91-like models of class A ($\alpha=1.5$ for the IMF). The various hatching
  styles indicate the origin of the gas. The initial BH mass is too
  small to be visible on these diagrams (dark gray hatching). The
  thick line is the mass of the central BH, as limited by Eddington
  luminosity. Our simulations were realized with 256\,000
  super-stars. Note that the ordinate mass units are different in each
  panel. For this top-heavy stellar spectrum, the role of stellar
  evolution is clearly dominant even in model A where the high stellar
  density boosts the collision rate. Panels (a) to (d) correspond to
  decreasing initial cluster mass (see text).}

  \label{fig:MBH_MCD91A}
\end{figure*}

\begin{figure*}
  \resizebox{\hsize}{!}{%
    \includegraphics{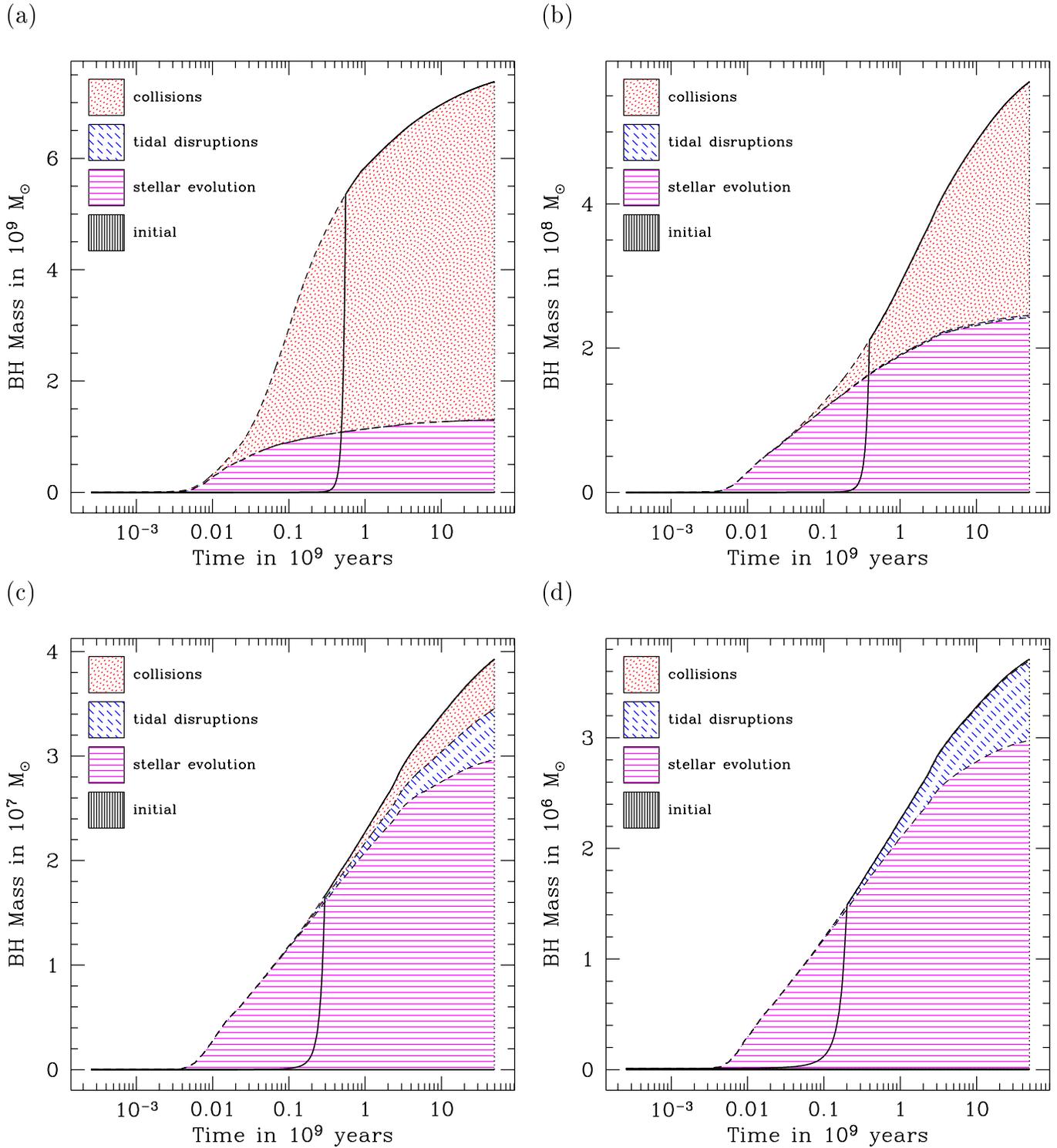}}

  \caption{ Same as Fig.~\ref{fig:MBH_MCD91A}, but for models of class
  B ($\alpha=2.5$).}

  \label{fig:MBH_MCD91B}
\end{figure*}

\begin{figure*}
  \resizebox{\hsize}{!}{%
    \includegraphics{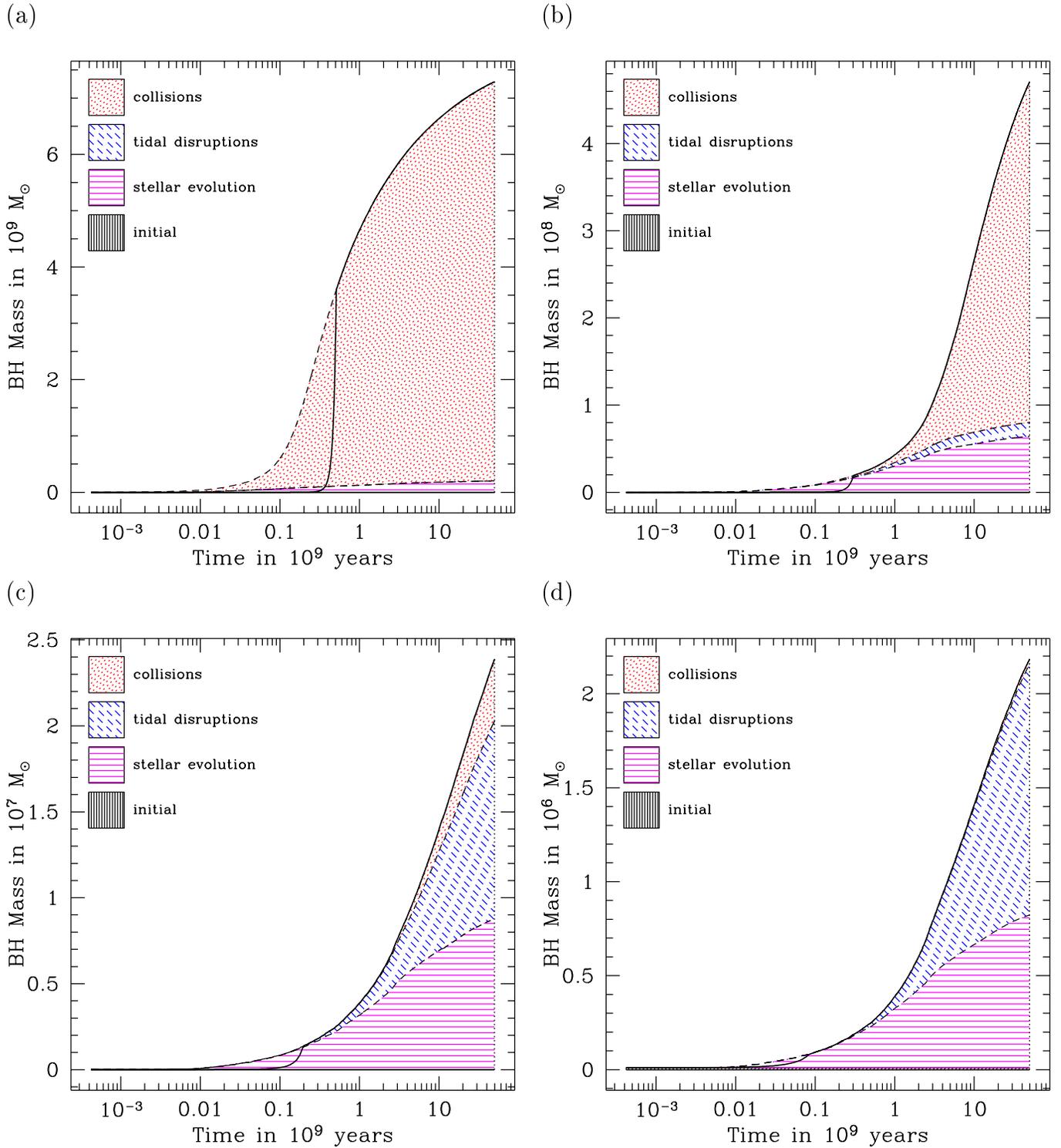}}

  \caption{ Same as Fig.~\ref{fig:MBH_MCD91A} and
  \ref{fig:MBH_MCD91B}, but for models of class C ($\alpha=3.5$). In
  this model with a stellar IMF strongly dominated by low masses, the
  role of stellar evolution is minimized so that collisions and tidal
  disruptions dominate the gas production rate.}

  \label{fig:MBH_MCD91C}
\end{figure*}
 
\begin{figure*}
  \resizebox{\hsize}{!}{%
    \includegraphics{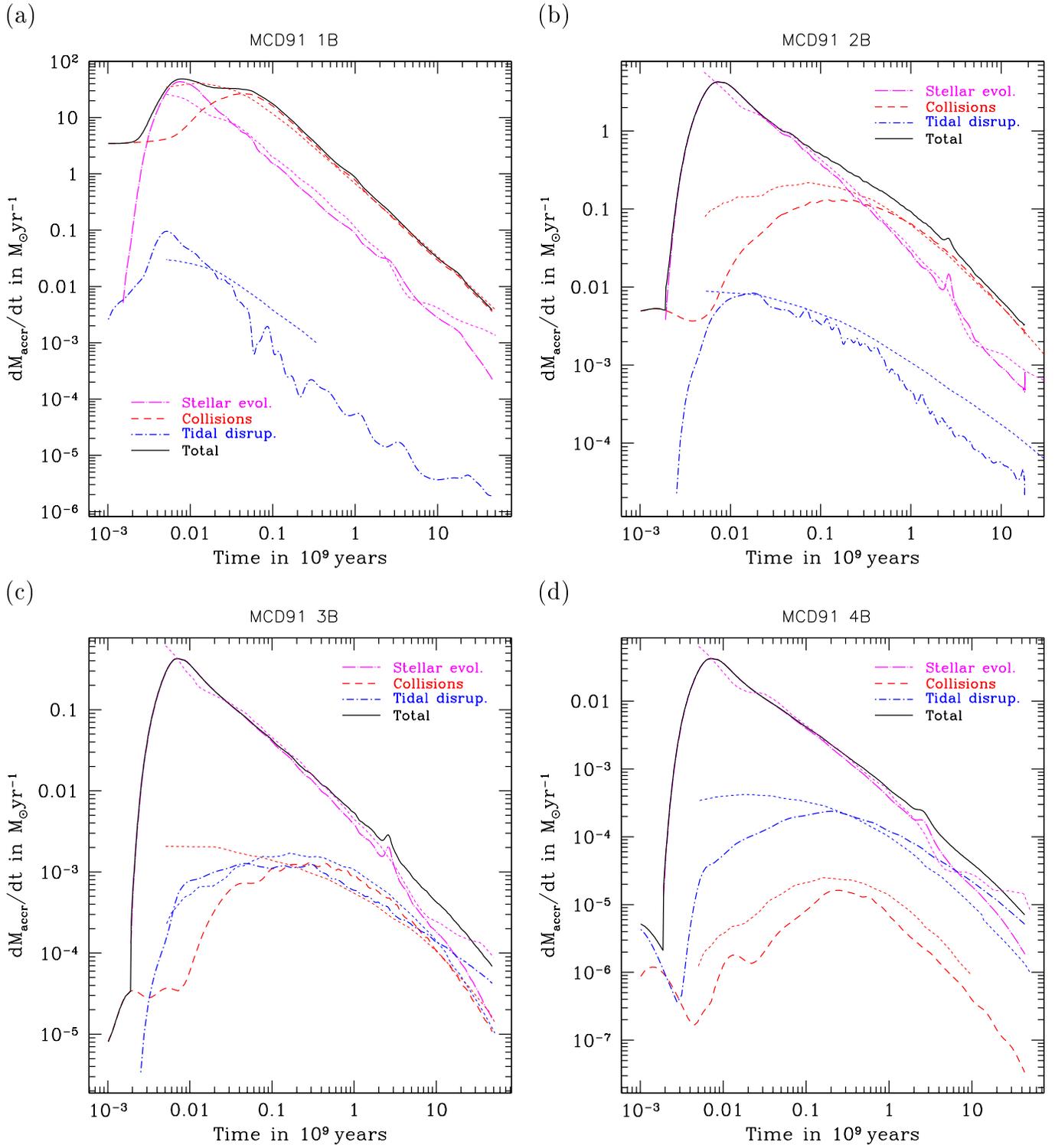}}

  \caption{ Evolution of the gas production rate for galactic nucleus
  models with initial conditions corresponding to models 1B--4B of
  MCD91. We plot the amount of gas the stars release per year through
  different channels: stellar evolution, collisions and tidal
  disruptions. Note that, at early times, only a fraction of this gas
  is accreted by the central BH while the remaining accumulates in
  some central reservoir. The thin dotted lines are the results of
  MCD91 but, for clarity, their total rates are omitted. Our
  simulations were realized with $10^6$ super-stars. The small-scale
  oscillations present in our curves are numerical noise. See text for
  further comments.}

  \label{fig:dMdt_MCD91B}
\end{figure*}

\begin{figure}
  \resizebox{\hsize}{!}{ \includegraphics{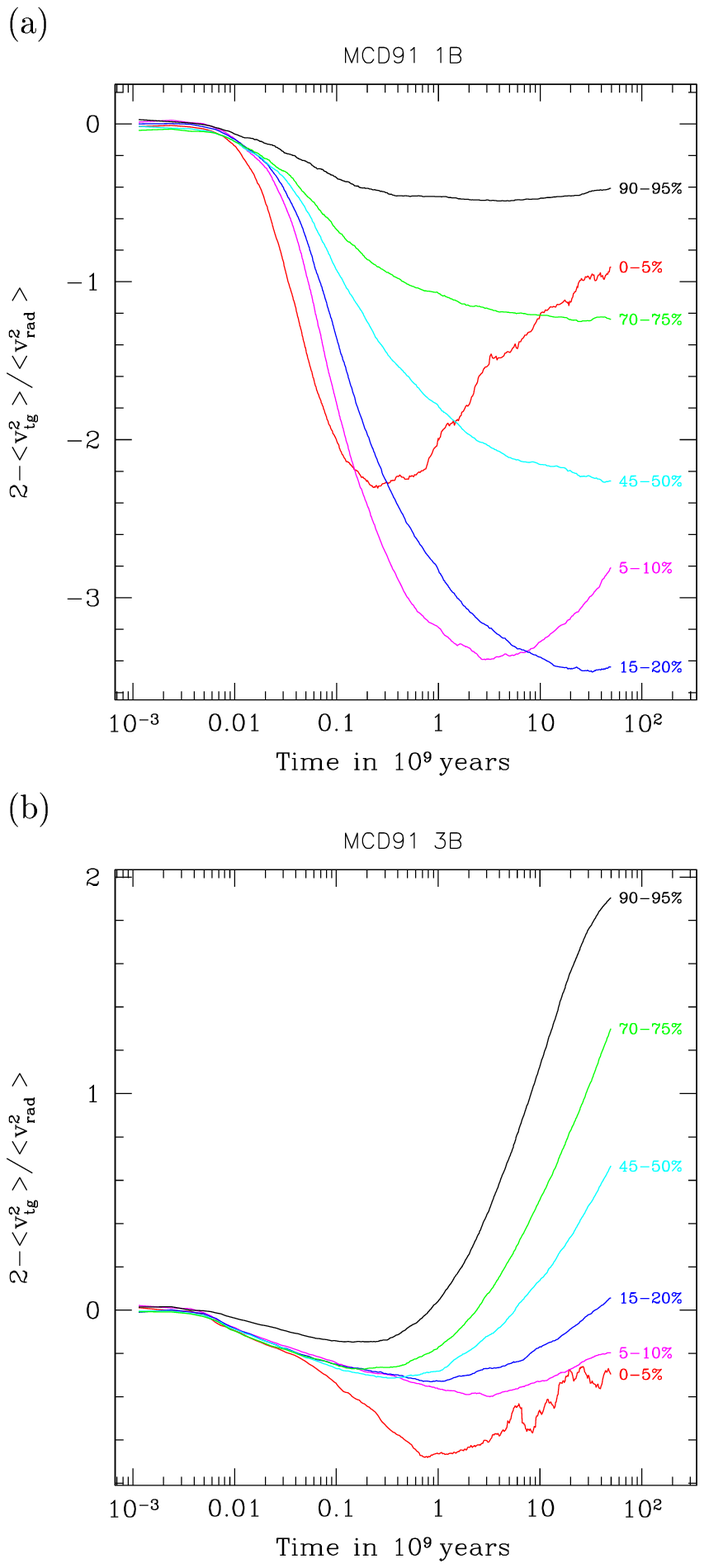} }

  \caption{ Evolution of the anisotropy for 2 models of class B
  simulated with $10^6$ super-stars. We show the anisotropy parameter
  averaged over Lagrangian shells bracketed by the indicated fraction
  of the (remaining) stellar mass. Note how strong a tangential
  anisotropy develops in model 1B, certainly in response to the
  adiabatic growth of the central BH. At later time, relaxation cause
  the central parts to slowly return to a more isotropic velocity
  distribution. The evolution of anisotropy in the lighter model 2B is
  strikingly different. For clarity, the curves have been slightly
  smoothed.}

  \label{fig:anis_MCD91}
\end{figure}

We have simulated all models specified by MCD91 with 256\,000
super-stars. For models of class B, we have redone the simulations
with $10^6$ super-stars. We basically mimic the initial conditions and
physics of MCD91. For instance, we used $\gamma=0.4$ for the Coulomb
logarithm. Note that MCD91's FP method imposes an isotropic velocity
distribution while our code allows anisotropy to develop. In addition
to the obvious differences imposed by the use of a very different
simulation algorithm, the following distinctions in the treatment of
the physics have to be noted:

\begin{itemize}
\item The collisions are treated much more realistically, on a
 particle-particle basis and outcomes are given by our SPH-generated
 grid for which realistic stellar structures have been used. The
 collisional modification of orbits is accounted for and mergers may
 occur.

\item The stellar evolution is slightly different from MCD91 (see Sec.~\ref{subsec:se}).

\item A ``continuous'' mass spectrum is used instead of the discrete mass 
classes of MCD91. To get the same average stellar mass as these
authors, the mass range is extended to $0.258 -
34.8\,M_{\odot}$. Also, masses as low as $0.01\,M_\odot$ may be
produced in collisions (smaller collisional products are not allowed)
while MCD91 use a ``hard'', constant minimum of $0.3\,M_\odot$.

\item We use a $M_{\ast}$--$R_{\ast}$ relation from MS stellar models 
\citep{SSMM92,MMSSC94,CDSBMMM99,CB00} to determine collisional cross-sections 
and tidal disruption radii.

\item Stellar evaporation, due to gradual energy gain through 2-body
relaxation (see Paper~I), is allowed in our models but MCD91
apparently enforce evolution at constant total mass which seems
reasonable because, for a cluster with no tidal truncation, diffusive
relaxation is expected to be inefficient. Indeed, it takes longer and
longer to increase the (negative) energy of a star toward $E>0$, as it
stays for a larger and larger fraction of its orbital time in
large-radius, low-density regions where relaxation is vanishingly
small \citep{Henon60b,Henon69}. For B models with $10^6$ super-stars,
we tried to forbid relaxation-driven stellar evaporation by discarding
``super-encounter'' that lead either super-star to escape the
system. The results appear not to be significantly altered by this
special treatment.

\end{itemize}

Among the results published by MCD91, those with which comparisons
are most easily carried out and which are of prime interest for us,
concern the growth of the central BH and how various processes
contribute to it. In Fig.~\ref{fig:end_MBH_MCD91}, for all 12 models
considered by MCD91, we compare the final mass of the central BH and
indicate which process contributed most to this mass. We confirm that,
unless the stellar mass spectrum is strongly bottom-heavy (case C,
$\alpha=3.5$) low-mass models are dominated by stellar evolution. C
models of low mass are the only ones for which tidal disruptions are a
significant fuel source. At higher (initial) stellar densities,
collisions dominate, with the densest A model as an exception. The main
source of discrepancy between our findings and those of MCD91 is the
more minor role of collisions in our simulations. While it is
difficult to evaluate how MCD91's use of fixed classes of $M_\ast$
translate in their collisional gas production rate, it is certain that
their $n=3$ polytropes models experience more mass loss in off-center
collisions than more realistic stars \citep{FB00c,FB02} and that their
mass-radius relation lead to an overall overestimate of collision
cross-section. A secondary source of mismatch is our different
prescription for stellar evolution. The temporal evolution of the
central mass (BH\,$+$\,gas reservoir) for all 12 models is depicted in
figures~\ref{fig:MBH_MCD91A} to \ref{fig:MBH_MCD91C}.

A more detailed comparison is realized for models of class B for which
MCD91 published the curves of the rate of gas production through each
process. Our results are reported on Fig~\ref{fig:dMdt_MCD91B}. Here
again, we notice that the main difference with MCD91 is that their
collisional rate is much higher at early times. This is probably due to
the presence of massive stars for which their assumptions about
stellar structure and radius should lead to the most severe
overestimate of collisional mass-loss. In fact, in regard of how
different (and more detailed) our treatment of collisions is, it is
very surprising how similar the collisional gas production rates are
at late times. The tidal disruption rates are very similar at early
time, with the exception of the least dense model (4B). At later
times, our tidal gas production rate decreases at a steeper rate (as
compared to MCD91) for the two densest models, while the contrary is
true for models 3B and 4B. One possible explanation for the lowest
late-time rates in dense models is that significant tangential
anisotropy develops in the central parts of these clusters, probably
in response to the rapid and, hence, nearly adiabatic, growth of the BH,
a process which does not significantly affect the ``light''
clusters. This aspect is illustrated in
Fig.~\ref{fig:anis_MCD91}. Obviously, stars on low eccentricity orbits
are less likely to enter the loss cone, an aspect of the dynamics that
MCD91 could not simulate with their isotropic code. On the other hand,
this does not explain why we get a higher late time disruption rate
for the lower density clusters.

\section{Conclusions.}
\label{sec:concl}

\subsection{Summary}

In this second paper about our Monte Carlo code for star cluster
simulations, we have described our inclusion of physical processes
pertaining to the dynamics of galactic nuclei.  

Taking advantage of the particle-based approach of the MC code,
collisions between MS stars are treated with a high level
of realism. The MC sampling reproduce the rate of collisions between
stars of various masses and the distribution of relative velocities
(and impact parameters) in a straightforward way. The outcome of
collisions are obtained by interpolation into a comprehensive database
of results from SPH simulations \citep{FB00c,FB02}. This is an
important improvement over previous works that included the role of
collisions in the dynamical evolution of galactic nuclei but relied on
simple-minded prescriptions for the results of collisions. In the
past, only
\citet{Rauch99} has attempted to use the outcome of a limited number of 
SPH simulations by M.~Davies to find fitting formulae for their
outcome and incorporate collisions in cluster models. It is, however,
doubtful that these results, obtained with polytropic stellar models
and from a relatively small domain of the parameter space can be
applied for realistic stars and other relative velocities and/or
impact parameters
\citep{FB00c}.

The second important feature of the dynamics of a galactic nucleus, as
compared to a globular cluster, is the likely presence of central BH
with a mass in excess of $10^6\,M_\odot$ (although some globular
clusters, like M~15, may harbor a central BH, see
\citealt{GPOCWH00} and \citealt{vdM01}). In our code, we assume the BH stays perfectly 
at the center (see below) and treat its contribution to the potential
as that of a Newtonian point mass. The neglect of relativistic effects
on stellar orbits is probably a good approximation, according to
\citet{Rauch99} who concluded that they seem to have no noticeable influence in
his simulations. The BH grows by accreting gas released by the stellar
system through stellar evolution, collisions and tidal
disruptions. Whole stars may also be swallowed if they directly plunge
through the horizon. This latter process completely supersedes tidal
disruption for MBH more massive than a few $10^8\,M_\odot$ because,
then, the tidal disruption radius is formally inside the horizon. For
the time being, the process of tidal disruption itself is treated as
simply as possible, by assuming complete disruption of every star that
enters the Roche zone around the BH. On the other hand, we test for
super-stars entering the so-called ``loss-cone'', i.e. getting onto
disruption orbits, in a detailed way by simulating the fine-grained
diffusion caused by relaxation on the direction of a super-star's
velocity.

Other improvements include a simple treatment of stellar evolution
which assumes that stars transform directly from MS to compact
remnants, in a similar spirit to what has been done by previous
investigators \citep{NS88,MCD91}. Also, we have implemented ``particle
doubling'' to maintain high resolution even in simulations where a lot
of stars are either destroyed or ejected from the cluster.

These new features have been extensively checked against
(semi-)analytical predictions and simulations from the literature. In
most cases, the tests are highly successful. In particular, collision
rates are nicely reproduced, not only when integrated over the whole
cluster but also as a function of distance from the center and of the
masses of stars. The effects on the stellar cluster of an
adiabatically growing central black hole are nearly perfectly in
agreement with theoretical predictions. The standard ``Bahcall and
Wolf'' $R^{-7/4}$ density cusp is obtained in the case tidal
disruptions are taken into account but collisions are switched off or
inefficient. In highly collisional models, a shallower cusp, with
exponent around $-0.5$ is produced, in good agreement with what was
reported in previous studies. Gas production by the stellar cluster
through various processes (tidal disruptions, collisions, stellar
evolutions) are also in good agreement with results from the
literature, obtained with a variety of numerical methods. Most of the
discrepancies can be easily explained. In particular, it appears that
the role of collisions has been overestimated in previous works, due
to over-simplified assumptions about the collisional outcome (complete
disruptions or simple semi-analytical treatment applied to polytropic
models) and, maybe, to their being included into the simulations in a
quite nonphysical way, in the case of direct Fokker-Planck
methods. Concerning tidal disruptions, some disagreement, for which we
have found no straightforward explanation, is observed with the works
of
\citet{DS82} and \citet{ASS01b}. 
These mismatches are not severe, however, and, as the resolution of
the simulations by \citet{DS82} was quite low\footnote{They used a few
thousands particles but their cloning algorithm increased the relative
resolution at large negative energies, i.e. close to the BH.} and the
results plotted by
\citet{ASS01b} come only from preliminary computations, we can not draw 
definitive conclusions from these comparisons. Furthermore, there is
no clear trend in these differences and we get better agreements in
other cases \citep[with, e.g., model~I of][]{DS83}, a fact which seems
to exclude any important flaw in our algorithm. Unfortunately,
$N$-body methods seem still a long way from allowing simulations of the
relaxational dynamics around a black hole and, thus, providing more
direct check of our approach and, more generally, of the applicability
of the loss-cone theory and the Chandrasekhar treatment of relaxation
in such a situation \citep[e.g.,][]{SK99}.

\subsection{Future work}
\label{subsec:futurework}

In Sec.~8.2 of paper~I, we have already mentioned many
improvements/additions that we plan to incorporate in future versions
of the code. Here, we update and complete this list:
\begin{enumerate}

\item {\bf Capture of compact stars by the central BH 
through emission of gravitational radiation}. This process has been
presented in Sec.~\ref{subsec:motiv}. Predicting the rate and
characteristics of these events has recently become a main focus of our
work and very encouraging results have already been reported in
\citet{Freitag01}.

\item {\bf Refined treatment of stellar evolution}. The most severe 
shortcoming of our present modeling of SE is the absence of giant
phase. Taking it into account should greatly enhance the number rates
of collisions and tidal stripping \citep{DBBS98,BD99,MT99,SU99}
although the amount of released gas may be limited due to the very low
density of giants' envelopes and this may not increase the BH's growth
as this gas would be liberated anyway through stellar
evolution. Others aspects of SE that we shall incorporate are:
progressive mass loss on the MS, natal kicks for neutron stars and collisional
rejuvenation. 

\item {\bf Refined treatment of tidal interactions}.
  
We should treat the hydrodynamical nature tidal disruptions with the
same level of realism that we achieved for collisions. This will be
essential if we want to cope with envelope-stripping of giant stars
\citep{DSGMG01} and other ``tidally perturbed'' stars \citep{AL01}. 
Stars can also be tidally captured by the central BH. As more and more
orbital energy is transfered to oscillations at each subsequent
pericenter passage, disruption is the most probable outcome
\citep{NPP92}.

Assuming that the BH stays fixed at the center of the cluster is an
over-simplification. If the central BH's wandering is of larger extent
than its tidal disruption radius $R_{\mathrm{disr}}$, there will be no
regime of empty loss cone
\citep{SR97}. For a cluster with core radius
$R_{\mathrm{c}}$, equipartition predicts a wandering radius of order
$R_{\mathrm{w}} \approx R_{\mathrm{c}}
\sqrt{M_{\ast}/M_{\mathrm{BH}}}$ \citep{BW76,LT80,CHL01}, and
\[
  \frac{R_{\mathrm{w}}}{R_{\mathrm{disr}}}\approx 400\, 
\left(\frac{R_{\mathrm{c}}}{1\,\mathrm{pc}}\right)
 \left(\frac{R_{\ast}}{R_{\sun}}\right)^{-1}
 \left(\frac{M_{\ast}}{M_{\sun}}\right)^{\frac{5}{6}}
 \left(\frac{M_{\mathrm{BH}}}{10^6\,M_{\sun}}\right)^{-\frac{5}{6}}.
\]
See \citet{MT99} for hints at the possible effects of the wandering on
the tidal disruption rate. \citet{Young77b} made a rough estimate of
the correction and deemed it not to alter the disruption rate
drastically. However, as suggested by \citet{AL01}, these motions of
the BH may allow stars that have been tidally perturbed to escape
further, disruptive, close interactions with the BH, which is of high
potential interest for the Galactic center.

\item {\bf Large angle scatterings}. 

2-body gravitational encounters with impact parameter of order or
smaller than $b_0 = G(M_1+M_2)/V_{\mathrm{rel}}^2$ lead to scattering
angles of order $\pi$.  Although they only contribute a fraction
$\ln(b_\mathrm{max}/b_0)^{-1}<0.1$ to the overall relaxation
\citep[ p.~198]{Henon73}, they may dominate the rate of evaporation 
from the cusp \citep{LT80,Goodman83} and of captures on relativistic
orbits \citep{SR97}. Such ``kicks'' can not be decomposed into smaller
deflections but can probably be introduced explicitly in
the MC code in a similar way as collisions.

\item {\bf Inclusion of binary stars}. 

In a ``normal'' population \citep{DM91}, most binaries have (internal)
orbital velocities smaller than the velocity dispersion near the
central BH (a few hundreds \kms) and will eventually be disrupted
through interactions with other stars. However, some small fraction
may be hard enough to survive and evolve into compact
binaries. Whether hard binaries will have an important dynamical role
has to be explored. Their interaction with the central BH is of
particular interest. Indeed, if it passes sufficiently close to the
BH, a binary will be tidally disrupted with the likely result of
ejecting one star out of the cluster at very high velocity and leaving
the other one bound to the BH \citep{Hills88,Hills91}. This is another
channel to form extreme mass ratio binaries to be detected by {\em
LISA}.

\item {\bf Interaction with a central accretion disk or gas cloud}.

The early evolution of galactic nuclei may well lead to the
accumulation of a quasi spherical central gas cloud with high enough a
density to interact strongly with the stellar cluster. This situation
has not yet been given the attention it deserves
\citep[see, however, ][ and references therein]{Spurzem92} but further 
investigations have been undertaken by Amaro-Seoane and collaborators
\citep{ASS01b,ASSJ01}.

In AGNs, stars may be captured by an accretion disk through repeated
impacts which can strongly reshape the stellar distribution in the
vicinity of the BH \citep{NS83,SCR91,Rauch95,VK98a,KS01,VC02}.  The
further stellar and orbital evolution of the disk-embedded stars is a
complex subject. Interesting possibilities include enhanced rate of
collisions and growth of massive stars by accretion of disk
material. Note that even if the interactions with the accretion disk
are not efficient enough to grind down orbits into the disk, stellar
formation probably occurs {\it in situ} \citep{Goodman02} so that the
presence of stars in the disk has to be expected anyway. A possible
way of accounting for the role of the accretion disk in numerical
models would be to use the MC code to simulate the outer
quasi-spherical parts of the cluster where relaxation is important and
couple it with a code like that of
\citet{Subr01} which treats the inner regions, where interactions with
the disk dominate the dynamics, in axisymmetrical geometry.

\item {\bf Gas dynamics}.

Including stellar evolution without a better prescription for the
fraction of gas that eventually finds its way to the central BH is
nearly pointless, as demonstrated by simulations in
\citet{MaThese}. Early studies
\citep{Bailey80,LF80,DDC87a,DDC87b,KLY87,NS88} concluded that most of the 
gas finds its way to the central BH but they lacked detailed account
of the feed-back on the gas of the energy released by the central
source and supernova explosions and of the complex, non-spherical,
evolving geometry of the gas flow \citep[see, e.g.,][ for recent
attempts at tackling these intricacies]{WBP99,CO01}. 

\end{enumerate}

This list can be lengthened virtually without end. But before we hurry
and include more and more complexity in our simulations, we must keep
in mind that each new process to be added comes with its own
uncertainties of both physical and numerical nature, so that the
impression of added ``realism'' may be misleading. In such a context,
it is all the more useful to dispose of a numerical tool flexible
enough to allow changes in the treatment of various physical effects
and fast enough to allow large sets of simulations to be conducted to
test for the influence of these modifications.

Another line along which we have to progress is to develop definite
observational predictions. Here are a few examples:
\begin{itemize}
  \item Surface luminosity and color profiles for central cusps.
  \item Rate and characteristics of radiation flares following the
    tidal disruption of a star.
  \item Appearance (and radial distribution) of stars modified by
    collisions or tidal interactions with the MBH.
  \item Rate and characteristics of gravitational waves signals from
    captured stars.
\end{itemize}

All examples but the first are complex problems of their own and have
already been the subject of many detailed, if not conclusive, studies.
Fortunately these aspects are essentially decoupled from the cluster
dynamics, in the sense that they have no obvious back-influence on it,
so that we should be able to ``map'' results from the literature on
the outcome of our simulations.

\begin{acknowledgements}
  
  Stellar models to initiate SPH collision simulations where kindly
  provided by the Geneva stellar evolution group, with precious help
  from Georges Meynet, and by Isabelle Baraffe and Corinne
  Charbonel. M.F. wants to thank Rainer Spurzem for interesting
  discussions and Gerald Quinlan and Roeland van der Marel for
  providing \texttt{pycode}, the BH adiabatic growth code. Comments by
  the anonymous referee helped to clarify the paper.

  Most simulations have been realized on the ``Beowulfs'' clusters GRAVITOR
  at Geneva
  Observatory\footnote{\texttt{http://obswww.unige.ch/$\sim$pfennige/gravitor/
  gravitor\_e.html}}, and ISIS at Bern University. This work has been
  supported by the Swiss National Science Foundation. The writing of
  this paper has been finished at Caltech, with partial support from
  NASA under grant NAG5-10707.

\end{acknowledgements}

\appendix

\section{Building of initial models of galactic nuclei}
\label{sec:initcond}

To obtain initial cluster realizations for our simulations, we proceed
in two stages: {\bf (1)} We set the radii $R_i$, specific kinetic
energies, $T_i$ and moduli of specific angular momentum, $J_i$ of all
super-stars\footnote{Remember that a super-star actually represents a
spherical shell of stars.} while trying to ensure dynamical
equilibrium. {\bf (2)} We set the stellar masses of the super-stars,
$M_i^{\ast}$, according to a given initial mass function (IMF). To get
an aged stellar population, we may also evolve this IMF according to
the ``ZAMS$\longrightarrow$remnant'' relation specified in
Sec.~\ref{subsec:se}. As the number of stars a super-star stands for
must be the same for all super-stars, this stage also implicitly
determines the super-star's mass,
$M_i=(N_{\ast}/N_{\mathrm{p}})M_i^{\ast}$ where $N_{\mathrm{p}}$
is the number of super-stars the model consists of and $N_{\ast}$ is
the number of stars represented by the model.

\subsection{Positions and velocities}

The safest way to obtain a system that is not only virialized
($2T_{\mathrm{cl}}+U_{\mathrm{cl}}=0$ where $T_{\mathrm{cl}}$ is the
total kinetic energy and $U_{\mathrm{cl}}$ the total gravitational
energy), but a genuine stationary solution of the collision-less
Boltzmann equation, is to start from a one-particle DF
$f(\vec{X},\vec{V})$ which depends on the position $\vec{X}$ and
velocity $\vec{V}$ only through isolating integrals of motions, namely
$E$ and $J$, for a stellar cluster that obeys spherical symmetry
\citep[][ Chap.~4]{BT87},
\begin{equation}
  f(\vec{X},\vec{V}) = F(E(\vec{X},\vec{V}),J(\vec{X},\vec{V})),
\end{equation}
with 
\begin{equation}
  E(\vec{X},\vec{V}) = \frac{1}{2} V^2 + \Phi(R) \mbox{~~and~~}
  J(\vec{X},\vec{V}) = RV_{\perp},
\end{equation}
where $R=|\vec{X}|$, $V=|\vec{V}|$, $V_{\perp}$ is the modulus of the
component of $\vec{V}$ perpendicular to $\vec{X}$ (with the cluster
center as origin of coordinates) and $\Phi$ is the (smooth)
gravitational potential. For the sake of simplicity, we only
considered initial cluster models with isotropic velocity
distributions for which $F$ is a function of $E$ only. Note that our
MC code can tackle any velocity distribution and that some level of
anisotropy develops during the run of most cluster simulations.

$\Phi$ is itself determined by the DF through 
Poisson equation:
\begin{equation}
  2\frac{\mathrm{d}\Phi}{\mathrm{d}R} +
  R\frac{\mathrm{d}^2\Phi}{\mathrm{d}R^2} = 4\pi G \rho(\Phi),
\end{equation}
with the density $\rho$ given by
\begin{equation}
  \rho(\Phi) = 4\pi \int_0^{\sqrt{-2\Phi}} 
  \mathrm{d}V\,V^2F(\frac{1}{2} V^2 + \Phi).
  \label{eq:rhoofphi}
\end{equation}

It is customary to define so-called relative energy and potential
through
\begin{equation}
  \Psi \stackrel{\mathrm{def}}{=} \Phi_0-\Phi \mbox{~~and~~}
  \varepsilon \stackrel{\mathrm{def}}{=} \Phi_0-E
\end{equation} 
with $\Phi_0$ chosen so that $F(\varepsilon)=0$ for $\varepsilon\le
0$. For a cluster of finite radius $R_{\mathrm{cl}}$,
$\Phi_0=-GM_{\mathrm{cl}}/R_{\mathrm{cl}}$.

Thus, to build a cluster model, we do the following:
\begin{itemize}
\item[{\bf (0)}] Choose an expression for $F(\varepsilon)$.
  Traditional choices are, among others, Plummer's or King's models
  \citep{BT87}.

\item[{\bf (1)}] Integrate $\Psi(R)$ and $M_r(R)$ with a
  Runge-Kutta scheme \citep{HNW87}:
\begin{equation}
  \frac{\mathrm{d}}{\mathrm{d}R}
  \left( \begin{array}{l}
      \Psi \\
      \Psi_{\mathrm{d}} \\
      M_r
    \end{array} \right) = 
  \left( \begin{array}{l}
      \Psi_{\mathrm{d}} \\
      -4\pi G\rho(\Psi) -\frac{2}{R}\Psi_{\mathrm{d}} \\
      4\pi\rho(\Psi)R^2
      \end{array} \right).
\label{eq:systED}
\end{equation}
Each evaluation of the function $\rho(\Psi)$ requires itself a
numerical integration of Eq.~\ref{eq:rhoofphi}. The integration of
system~\ref{eq:systED} is terminated either when the relative potential
reaches 0 (for tidally truncated models) or when $M_r$ has attained some
asymptotic value. At that point, we have obtained array
representations of $R$, $\Psi$, $\rho$ and $M_r$.  We re-normalize
them to the ``$N$-body'' system of units (see Sec.~\ref{sec:units}).
\item[{\bf (2)}] For each super-star, radius $R_i$ is randomly selected
  according to the probability density $\mathrm{d}M_r/\mathrm{d}R$.
  This is done by creating a random number $X_{\mathrm{ran}}$ with
  uniform probability over $[0;1[$ and (numerically) inverting the
  $M_r(R)$ relation: $R_i=M_r^{-1}(X_{\mathrm{ran}})$.
\item[{\bf (3)}] Once the radius $R_i$ of super-star $i$ is determined,
  we have to select a velocity $V_i$ according to distribution
  $g(V)\propto V^2 F(\frac{1}{2} V^2 + \Phi(R_i))$. Here we use a
  simple rejection method \citep[][ Sec.~7.3]{PTVF92} with a constant
  upper bound given by $-2 \Phi(R_i) F(\Phi(R_i))$.\footnote{Bound
    particles have $V^2/2+\Phi(R)<0$. Furthermore, well-behaved DF have
    $\mathrm{d}F/\mathrm{d}E<0$ so that the maximum value at a given
    $R$ is $F(\Phi(R))$.} The specific kinetic energy of the super-star
  is thus $T_i=V_i^2/2$. To set the specific angular momentum $J_i$ 
  with account of isotropy, we generate another random number
  $X_{\mathrm{ran}}$ and compute 
  $V_{\mathrm{rad}}=V_i(1-2X_{\mathrm{ran}})$ and
  $J_i=R_i\sqrt{V_i^2-V_{\mathrm{rad}}^2}$.
  
\item[{\bf (4)}] Finally, perfect virial energy balance is enforced by
  a slight re-scaling of the velocities.
\end{itemize} 

In its present form, this procedure does not explicitly allow for a
central BH.  But if we add such a point mass at the center with a very
small mass (as compared to $M_{\mathrm{cl}}$), it will only slightly
perturb the potential energies of the innermost super-stars and the
resulting system will still be very close to dynamical equilibrium.
This is the reason why we must always start simulations with ``seed''
black holes instead of already grown (super-)massive ones. An
advantage of this method is that the integrated influence of the BH's
growth on the stellar system is ``automatically'' computed! The main
drawback is that we cannot start with models that represent today's
galactic nuclei but have to guess initial conditions that lead to such
configurations after a Hubble time. This has not yet been explored
systematically.

The cluster produced with this algorithm has no mass spectrum, i.e.
all super-stars have the same mass
$M_{\mathrm{p}}=M_{\mathrm{cl}}/N_{\mathrm{p}}$. We now explain
how we construct a stellar mass spectrum.

\subsection{Masses}

We model IMFs that are piece-wise power-laws,
\begin{equation}
  \frac{\mathrm{d}N_{\ast}}{\mathrm{d}M_{\ast}} \propto
  M_{\ast}^{-\alpha_k} \mbox{~~for~~} M_{k-1} \le M_{\ast}\le M_k,
\end{equation}
between some $M_0=M_{\mathrm{min}}$ and $M_K=M_{\mathrm{max}}$.

For a given set of $M_k$ ($k=0,\ldots,K$) and
$\alpha_k$ ($k=1,\ldots,K$). The \emph{un-normalized} number of stars
with masses $\le M_{\ast}$ is, for $M_{k-1} \le M_{\ast}\le M_k$:
\begin{eqnarray}
  N(M_{\ast}) &=& N_{k-1} + C_k\int_{M_{k-1}}^{M_{\ast}}
  \frac{\mathrm{d}N_{\ast}}{\mathrm{d}M_{\ast}} \mathrm{d}M_{\ast}
    \nonumber \\
    &=& N_{k-1} +
    \frac{1}{1-\alpha_k}\left(M_{\ast}^{1-\alpha_k}-M_{k-1}^{1-\alpha_k}\right)
\end{eqnarray} 
with $C_k=C_{k-1}M_{k-1}^{(\alpha_k-\alpha_{k-1})}$ (we can set
$C_1=1$). Once the $N_k$ have been computed, we randomly determine the
stellar mass of each super-star in turn. We first generate a random
number $N_{\mathrm{ran}}$ with uniform $[0;N_K]$ distribution ($N_K$
is the un-normalized total number). We then find index $j$ such that $
N_{j-1} \le N_{\mathrm{ran}} \le N_j$ and invert $N(M_\ast)$ to find
the stellar mass for super-star $i$:
\begin{equation}
  M_i^{\ast} = \left( M_{j-1}^{1-\alpha_j} + 
    (1-\alpha_j)\frac{N_{\mathrm{ran}}-N_{j-1}}{C_j} 
  \right)^{\frac{1}{1-\alpha_j}}.
\end{equation}

Note that we never need to state the actual total number of stars
(or, equivalently, the total mass in $M_{\sun}$) or the  size of
the cluster in pc when building initial models. This must only be
specified before starting an evolutionary Monte Carlo simulation as
these mass and size scales determine the relative importances of various
processes (e.g. relaxation vs. collisions) and allows to translate the 
$N$-body time units into years.

\label{app:CI}

\bibliographystyle{apj} 

\bibliography{aamnem99,biblio} 

\end{document}